\documentclass[a4paper,12pt]{article}
\pdfoutput=1
\usepackage{jheppub}  
\usepackage{amsmath,amssymb}
\usepackage{graphicx}
\usepackage{float}
\usepackage[export]{adjustbox}
\usepackage[utf8]{inputenc}
\usepackage{slashed}
\usepackage{bbold}
\usepackage{physics}
\usepackage{mathtools}
\numberwithin{equation}{section}
\usepackage{xcolor}

\def\pol{\epsilon}
\def\nn {\nonumber}
\author[a]{Charalampos Anastasiou}
\affiliation[a]{Institute for Theoretical Physics, ETH Zurich, 8093
  Z\"urich, Switzerland}
\emailAdd{babis@phys.ethz.ch}

\author[a]{Julia Karlen}
\emailAdd{karlenj@phys.ethz.ch}
 
\author[b]{George Sterman}
\affiliation[b]{ C.N.\ Yang Institute for Theoretical Physics and Department of Physics and Astronomy\\
Stony Brook University, Stony Brook NY, 11794-3840 USA}
\emailAdd{george.sterman@stonybrook.edu}

\author[b]{Aniruddha Venkata} 
\emailAdd{aniruddha.venkata@stonybrook.edu}
\title{Locally finite two-loop amplitudes for electroweak production through gluon fusion}

\abstract{
The computation of two-loop amplitudes for the production of multiple Higgs and electroweak gauge bosons via gluon fusion with exact dependence on quark masses relies primarily on numerical methods. 
We propose a framework that enables their numerical evaluation in momentum space. The method is inspired by the factorization of infrared divergences in QCD scattering amplitudes. It extends techniques introduced for electroweak gauge boson production from quark-antiquark annihilation to processes with external gluons. 
By combining diagrammatic integrands, we make use of local cancellations between diagrams that automatically eliminate most non-factoring infrared singularities. With a limited number of counterterms, we then derive two-loop integrands for which all soft and collinear singularities factorize locally. 
We hope that the local subtraction techniques presented in this article will play a useful role in extending the local factorization formalism to two-loop amplitudes for arbitrary processes.
}

\begin{document}
\maketitle

\section{Introduction}

Increasing the precision of collider phenomenology requires improved numerical control over amplitudes for the production of multiple massive particles from the full range of partonic initial states. Here we adapt methods introduced for electroweak gauge boson production from quark-antiquark annihilation in \cite{Anastasiou:2020sdt,Anastasiou:2022eym} to processes with external gluons. This approach exploits the factorization properties of hard-scattering amplitudes to organize sets of infrared subtractions.

Amplitudes for scattering processes with large momentum transfers to all outgoing lines factorize into a soft function, jet functions, and a hard function \cite{Sen:1982bt, Catani:1998bh, Sterman:2002qn, Dixon:2008gr, Collins:2011zzd, Feige:2014wja, Erdogan:2014gha, Ma:2019hjq}. The soft and jet functions are infrared singular in four dimensions and contain all long-distance information on overall color flow and the velocities of external particles. The hard function summarizes all remaining information, from interactions at short distances and process-dependence. This factorization of scattering amplitudes holds at leading power to all orders in perturbation theory.

The universality of the soft and jet functions motivates the development of methods where the infrared-divergent parts of an amplitude for a process with many external particles are assembled from known analytic computations of amplitudes for simpler processes. The process-dependent hard factor, which is finite in $D=4$ dimensions, can in principle be computed numerically. To enable this final step, however, amplitude factorization should be made manifestly local in momentum space, generating an integrand for the hard function that is integrable in infrared limits and convergent in the ultraviolet. If these conditions are met, the hard function can be integrated directly in momentum space, using numerical methods (see, for example, Refs.~\cite{Soper:1999xk,Nagy:2006xy,Gong:2008ww,Becker:2010ng,Assadsolimani:2009cz,Becker:2012aqa,Becker:2011vg,Becker:2012bi,Gnendiger:2017pys,Seth:2016hmv,Capatti:2019edf,Capatti:2020ytd,Capatti:2020xjc,TorresBobadilla:2020ekr,Kermanschah:2021wbk,Rios-Sanchez:2024xtv}). 

Ref.~\cite{Anastasiou:2018rib} presented, as a ``proof of concept", examples of momentum-space subtractions for characteristic two-loop master integrals with infrared singularities. In Ref. \cite{Anastasiou:2020sdt}, progressing from sample integrals to the full set of amplitudes for specific gauge theory processes, it was shown that singularities can indeed be factorized locally at two loops, in QED amplitudes for electron-positron annihilation into off-shell photons. 
 In Ref. \cite{Anastasiou:2022eym}, the local factorization of infrared singularities was demonstrated in two-loop QCD amplitudes for quark pair annihilation into heavy colorless final states. This factorization is carried out by a limited number of infrared counterterms, fewer than the number of diagrams, and many fewer than the numbers of infrared-singular momentum regions in the full set of diagrams contributing to the process. 
It does so by organizing momentum flows to take advantage of cancellations between diagrams that are implied by factorization, and occur locally in momentum space.

A natural next problem to tackle is the extension of the local factorization method of Refs.\ \cite{Anastasiou:2020sdt,Anastasiou:2022eym} to two-loop amplitudes with initial state gluons. In this article, we make progress in this direction, focusing specifically on the production of colorless final states in gluon fusion processes at two loops. We construct amplitude integrands in which soft and collinear singularities factorize locally. In the processes that we consider, a set of heavy color-neutral particles are produced by the incoming gluons through the mediation of a heavy quark loop. We assume that the final state is not at the threshold value of the production for a pair of heavy quarks. Then, at two loops only a single loop momentum can generate infrared singularities, while the heavy quark loop integral remains infrared finite.

The factorization of collinear singularities is a consequence of gauge invariance, and is demonstrated by the application of Ward identities. In our factorization-based construction of local infrared counterterms for two-loop gluon-fusion amplitudes, we encounter certain features in common with the two-loop amplitudes of electroweak production from quark annihilation~\cite{Anastasiou:2022eym} and the analogous $e^+ e^-$ annihilation processes in QED~\cite{Anastasiou:2020sdt}. The essential challenge is to find an appropriate implementation of the relevant Ward identities.

We recall that Ward identities are not satisfied in a completely local fashion in momentum space, and famously require shifts in loop momenta.
At the same time, when loop momenta are assigned in a coordinated fashion among diagrams, almost all non-factorizing infrared singularities cancel at the level of integrands, leaving simple remainders, two terms that differ only in their signs and in the values of loop momentum. These differences (called {\it shift mismatches} in Ref.~\cite{Anastasiou:2022eym}) clearly integrate to zero. Shift mismatches would nevertheless spoil factorization locally, and we need to eliminate them to achieve numerical integrability. As in Ref.~\cite{Anastasiou:2022eym}, we will introduce a set of counterterms that integrate to zero and are engineered to cancel all shift mismatches. In this way, a limited number of counterterms implements local cancellation for the remaining unphysical singularities, which normally cancel only after integration. Physical infrared singularities, which factorize, are also subtracted locally, by a universal one-loop scalar form factor amplitude times the infrared finite Born amplitude. 

We begin our analysis in Sec.~\ref{sec:framework_notation} with a brief review of the results of Ref.\ \cite{Anastasiou:2022eym}, which treated quark pair annihilation, then introduce the notation and analyze the two-loop structure of the integrands for gluon-gluon fusion to multi-Higgs final states. Additionally we introduce a decomposition of the triple-gluon vertex, which facilitates the organization and analysis of the infrared behavior of all diagrams, discussed in Sec.~\ref{sec:IRsing_channels}. In Sec.~\ref{sec:factorization}, we demonstrate the factorization of the collinear limits of the amplitude integrands, for which a suitable loop momentum routing is necessary. We continue in Sec.~\ref{sec:NonLocalSings} with the construction of shift counterterms to remove non-local cancellations of remaining non-factorizable singularities in collinear limits. These singularities, which cancel among themselves after integration, can be made to cancel locally after the addition of appropriate shift counterterms. In the terminology we introduce, they are described as ``shift integrable". In Sec.~\ref{sec:UVCT}, we proceed to discuss ultraviolet singularities and the construction of appropriate counterterms analogous to Refs.~\cite{Anastasiou:2020sdt,Anastasiou:2022eym}. Finally we discuss the analytic and numerical checks carried out to verify the infrared and ultraviolet finiteness of the hard function in Sec.~\ref{sec:numcheck}. We conclude with a brief summary of our results.

\section{Framework and notation}
\label{sec:framework_notation}

We begin our discussion with a review of the results of Ref.~\cite{Anastasiou:2022eym}, introducing the framework for quark-antiquark annihilation to colorless electroweak final states. We then adapt this framework to gluon-gluon fusion to colorless states.

\subsection{Quark pair annihilation at two loops}

Reference~\cite{Anastasiou:2022eym} discussed processes of the form $q\bar q \to ew$, where $ew\, \equiv\, \{q_1 \dots q_n\}$ denotes any final state with only colorless particles, with non-lightlike momenta $q_i^\mu$. Always working here at tree level in the electroweak interactions, the perturbative expansion of the corresponding QCD amplitudes takes the form,
\begin{align}
    M_{q\bar q \to ew} =&    {\cal M}_{q\bar q \to ew}^{(0)} + \int \frac{d^D l}{\left(2 \pi \right)^D} {\cal M}_{q\bar q \to ew}^{(1)}\left( l \right) + \,  \int \frac{d^D l}{\left(2 \pi \right)^D}\frac{d^D k}{\left(2 \pi \right)^D} {\cal M}_{q\bar q \to ew}^{(2)}\left( k, l \right) + \ldots\ . \nn\\
\end{align}
Here and in the following, the notation ${\cal M}^{(m)}$ denotes the integrand of the $m$-loop amplitude, including external spinors, generated directly from the QCD Lagrangian density in Feynman gauge.

Introducing a slight variation of notation from Ref.~\cite{Anastasiou:2022eym}\footnote{Compared to Ref.~\cite{Anastasiou:2022eym}, we introduce the notation ${\widehat {\cal M}}$ for the two-loop integrand subtracted for non-factorizing singularities that cancel, as described below. We denote below these subtractions as $\Delta{\cal M}$. In addition, we use the symbol ${\cal H}$ for the fully infrared and ultraviolet finite integrand, in place of ${\cal M}_{\mathrm{finite}}$.},
the infrared and ultraviolet finite, process-dependent integrand amenable to numerical evaluation is found by iterative subtractions involving form factor integrands, 
\begin{eqnarray}
\label{eq:EWKsubtraction1}
{\cal H}_{q\bar q \to ew}^{(1),R}(k) &=& {\cal M}_{q\bar q \to ew}^{(1),R}(k) - {\cal F}_{q\bar q}^{(1),R}(k) \left[\mathbf{P}_1 \widetilde{\cal M}_{q\bar q \to ew}^{(0)} \mathbf{P}_1 \right] ,\\[2mm]
 \label{eq:EWKsubtraction2}
  {\cal H}_{q\bar q \to ew}^{(2),R}(k,l) &=& {\widehat {\cal M}}_{q\bar q \to ew}^{(2),R}(k,l) - {\cal F}_{q\bar q}^{(2),R}(k,l) \left[\mathbf{P}_1  \widetilde{\cal M}_{q\bar q \to ew}^{(0)} \mathbf{P}_1 \right]
  \nonumber \\
  &\ & \hspace{5mm}
  - {\cal F}_{q\bar q}^{(1),R}(k) \left[ \mathbf{P}_1 \widetilde{\cal H}_{q\bar q \to ew}^{(1),R}(l) \mathbf{P}_1 \right],
\end{eqnarray}
where ${\cal H}_{q\bar q \to ew}^{(m)}$ denotes the $m$-loop hard finite remainder integrand and ${\cal F}_{q\bar q}^{(m)}$ denotes the $m$-loop form factor amplitude integrand, both including external Dirac spinors. 
The superscript $R$ indicates that the terms are UV regularized by constructing and subtracting local UV counterterms, whose construction follows the method described in Sec.~\ref{sec:UVCT}. The factor $\mathbf{P}_1$ is defined as 
\begin{eqnarray}
    \mathbf{P}_1 &\equiv& \frac{\slashed{p}_1\slashed{p}_2}{2\,p_1\cdot p_2}\, ,
    \label{eq:p_1-def}
\end{eqnarray}
which projects on $p_1$ solutions $u(p_1,\lambda)$ when acting from the left, and $p_2$ solutions from the right,
\begin{eqnarray}
    \mathbf{P}_1\, u(p_1,\lambda) &=& u(p_1,\lambda),
    \nn \\
    [2mm]  \bar v(p_2,\sigma)\, \mathbf{P}_1 &=& \bar v(p_2,\sigma)\, .
\end{eqnarray}
These projections isolate terms with singularities at the next loop order, as constructed in Ref.~\cite{Anastasiou:2020sdt}. They enclose the Born amplitude and the one-loop hard amplitude with spinors stripped, denoted $\widetilde{\cal M}_{q\bar q \to ew}^{(0)}$ and $\widetilde{\cal H}_{q\bar q \to ew}^{(1),R}(l)$ respectively. As a result, $\widetilde{\cal M}_{q\bar q \to ew}^{(0)}$ and $\widetilde{\cal H}_{q\bar q \to ew}^{(1),R}(l)$ have two spinor and two color indices, which we do not exhibit. In particular, the spinor indices of ${\cal F}$ contract with the outer indices of ${\bf P}_1$~\footnote{See  Eq. (3.25) of Ref.~\cite{Anastasiou:2020sdt}.}. The Dirac structure can be non-trivial because we need not neglect the masses or momenta of particles in the short-distance functions. On the other hand,  
for processes with a colorless final state, as we consider in this article, the color indices are always of the form $\delta_{ab}$ in the relevant color representation defined by the initial state.

Our ability to choose identical projectors for particle and antiparticle in Eq.~\eqref{eq:p_1-def} is a feature of the two incoming spinors. More generally, anticipating more than two external partons, the projectors for $p_1$ and $p_2$ could be chosen independently, in terms of solutions to the Dirac equation,
\begin{eqnarray}
   \mathbf{P}_1 &=& \frac{1}{2p_1\cdot \bar \eta_1}\, \sum_\lambda u(p_1,\lambda) u^\dagger(p_1,\lambda),
    \nonumber\\[2mm]
  \mathbf{P}_2  &=& \frac{1}{2p_2\cdot \bar \eta_2}\, \sum_\lambda \gamma^0 v(p_2,\lambda) \bar v(p_2,\lambda)\, ,
\end{eqnarray}
where we parameterize the incoming gluon momenta as $p_i^\mu = (p_i\cdot \bar\eta_i) \, \eta_i^\mu $, in terms of lightlike ``velocity" vectors $\eta_i^\mu$ and their parity-reversed partners, $\bar \eta_i^\mu$, with $\eta_i\cdot \bar \eta_i=1$.
 The equality of these two expressions when $\eta_2=\bar \eta_1$ follows from the relation $\gamma^0 u(p_i,\lambda)=u(\bar p_i,\lambda)$.

The two-loop integrand on the right-hand side of Eq.~\eqref{eq:EWKsubtraction2} is defined in terms of the standard integrand by the subtraction of counterterms,
\begin{eqnarray}
    {\widehat {\cal M}}_{q\bar q \to ew}^{(2),R}(k,l) = {\cal M}_{q\bar q \to ew}^{(2),R}(k,l) - 
    \Delta {\cal M}_{q\bar q \to ew}^{(2),R     }(k,l)\, .
    \label{eq:bfMcalMDeltaM}
\end{eqnarray}
These counterterms do not affect the final amplitude, because they are constructed to integrate to zero,
\begin{eqnarray}
    \int d^D k\, \int d^D l \, \Delta {\cal M}_{q\bar q \to ew}^{(2),R}(k,l) = 0\, .
\end{eqnarray}
The infrared counterterms $\Delta {\cal M}_{q\bar q \to ew}^{(2),R}(k,l)$ in Eq.~\eqref{eq:bfMcalMDeltaM} are of two types in Ref.~\cite{Anastasiou:2022eym}, where they are referred to as ``loop polarization" and ``shift" IR counterterms. Of these, the former are necessary to factor subleading corrections at two loops, canceling 
non-longitudinal, leading power polarizations of a virtual gluon that connects a jet loop correction (momentum $k$ in Eq.~\eqref{eq:EWKsubtraction2}) to the tree-level hard subdiagram. The ``shift" counterterms eliminate the need for changes of variables (shifts, for momentum $l$ in Eq.~\eqref{eq:EWKsubtraction2}) in one-loop hard subdiagrams that are necessary to implement Ward identities locally in the standard formulation of perturbation theory.

The essential feature of Eqs.~\eqref{eq:EWKsubtraction1} and \eqref{eq:EWKsubtraction2} is that they apply at the level of the integrand. All infrared
dependence is generated by the integrations of the factors ${\cal F}^{(1)}(k)$ and ${\cal F}^{(2)}(k,l)$. This is an
extension of normal factorization. As we have seen, it requires modifications of the graphical integrand
at two loops, ${\cal M}_{q\bar q \to ew}^{(2),R}$, relative to the direct application of
Feynman rules, replacing it with 
a different function,
${\widehat {\cal M}}_{q\bar q \to ew}^{(2),R}$, whose integral is the same as that of the original integrand. These modifications can be implemented by the addition of infrared counterterms (and UV subtractions), as indicated in Eq.~\eqref{eq:bfMcalMDeltaM}, which vanish upon integration but make Ward identities local. An important feature of the construction is that the number of counterterms grows no faster than the number of diagrams. This is made possible by organizing the calculation to take advantage of cancellations that occur automatically between diagrams that are related by gauge invariance.

By employing Eqs.~\eqref{eq:EWKsubtraction1} and \eqref{eq:EWKsubtraction2}, all one-loop and two-loop amplitudes for the production of heavy electroweak gauge bosons in quark annihilation processes were rendered locally finite and integrable in $D=4$ dimensions.

\subsection{Gluon fusion to colorless final states at two loops}

In this paper, we treat by the same method amplitudes for gluon fusion to colorless, electroweak final states, through a heavy quark loop.  As noted above, we assume kinematics that avoids heavy quark pair thresholds. The one-loop amplitudes are at the lowest perturbative order and have no infrared singularities~\footnote{However, they may require locally an ultraviolet counterterm which is discussed in Section~\ref{sec:UVCT}.}.

For the two-loop amplitudes, the aim again is to generate an infrared finite, process-dependent hard scattering function amenable to numerical evaluation by the addition of a manageable number of counterterms. Compared to Ref.~\cite{Anastasiou:2020sdt}, we will again find it necessary to add IR shift counterterms in the definition of our two-loop integrand, but not loop polarization counterterms, which only appear at two loops in the IR part. 
 
The subtraction takes the form, analogous to Eqs.~\eqref{eq:EWKsubtraction1} and \eqref{eq:EWKsubtraction2} for pair annihilation,
\begin{eqnarray}
\label{eq:GGsubtraction-full}
{\cal H}_{gg \to {\rm colorless}}^{(2),R}(k,l) &=& {\widehat {\cal M}}_{gg \to {\rm colorless}}^{(2),R}(k,l) - {\cal F}_{gg}^{(1),R}(k) \, 
\left[\mathbf{P}^g_2\,
{ \widetilde{\cal M}}_{gg \to {\rm colorless}}^{(1),R}(l)\,
\mathbf{P}^g_1\right]\,.
\nn\\
\end{eqnarray}
As in the fermion case above, we suppress the spin (vector) indices of the matrices $\mathbf{P}^g_i$, which are constructed to project on the physical polarizations of the incoming gluons,
\begin{eqnarray}
\mathbf{P}_i^g\cdot \epsilon(p_i) = \epsilon(p_i)\, ,
\end{eqnarray}
for $i=1,2$. For gluons, $\mathbf{P}^g_i$ is a symmetric matrix. In the back-to-back frame, the two projectors can both be chosen to be equal to the transverse metric, $\eta_\perp^{\mu\nu}$, analogous to Eq.~\eqref{eq:p_1-def} for a fermion pair.
Also by analogy to Eq.~\eqref{eq:EWKsubtraction1} for fermions, quantities with tildes, like ${ \widetilde{\cal M}}_{gg \to {\rm colorless}}^{(1),R}(l)$, are stripped of external polarizations, including their color. They thus have two free vector and color indices. The colorless nature of the final state ensures that the color tensor is always $\delta_{ab}$. We will exhibit these color indices only when necessary for the argument. Notice that in Eq.~\eqref{eq:GGsubtraction-full}, external polarizations are included in the form factor ${\cal F}_{gg}^{(1),R}(k)$ and the resulting integrable function ${\cal H}_{gg \to {\rm colorless}}^{(2),R}(k,l)$.

The explicit two-loop calculations in the subsequent sections of this article, 
show that we can simplify Eq.~\eqref{eq:GGsubtraction-full} to a form in which the external polarizations act directly on the one-loop IR finite amplitude, while all infrared singularities are absorbed into a one-loop scalar function, denoted ${\cal F}_{\rm scalar}$,
\begin{eqnarray}
\label{eq:GGsubtraction}
{\cal H}_{gg \to {\rm colorless}}^{(2),R}(k,l) &=& {\widehat {\cal M}}_{gg \to {\rm colorless}}^{(2),R}(k,l) - {\cal F}_{\rm scalar}^{(1),R}(k) \,  { {\cal M}}_{gg \to {\rm colorless}}^{(1),R}(l)\, .
%\nn \\
\end{eqnarray}
We do not anticipate such a simplification at higher order. Beyond two loops we expect a result analogous to Eq.~\eqref{eq:EWKsubtraction2}, which includes spin non-diagonal contributions.
In Eq.~\eqref{eq:GGsubtraction}, the two-loop integrand ${\widehat {\cal M}}_{gg \to {\rm colorless}}^{(2),R}$ will include a shift counterterm through its definition
\begin{eqnarray}
{\widehat {\cal M}}_{gg \to {\rm colorless}}^{(2),R}(k,l)
&=&
{\cal M}_{gg \to {\rm colorless}}^{(2),R}(k,l) - \Delta {\cal M}_{gg \to {\rm colorless}}^{(2),R}(k,l)\, .
\label{eq:bf-M-def}
\end{eqnarray}
The subtraction $\Delta {\cal M}^{(2)}$ either has the effect of shifting loop momenta in certain contributions to ${\cal M}^{(2)}$ to make necessary Ward identities local, or is proportional to a set of terms that vanish identically by a Ward identity. In either case, all contributions to $\Delta {\cal M}^{(2),R}_{gg \to {\rm colorless}}$ vanish after (and only after) integration over one integral,
\begin{eqnarray}
    \int d^Dl\, \Delta {\cal M}_{gg \to {\rm colorless}}^{(2),R}(k,l)=0\, .
    \label{eq:bf-Delta-M-vanishes}
\end{eqnarray}
We will discuss the construction of $\Delta {\cal M}$ in detail below, but emphasize again that in amplitudes for $gg\to {\rm colorless}$ through a heavy quark loop away from quark pair thresholds, the infrared singularities factor from the quark loop, and the hard scattering begins at one loop. For such processes, $\Delta {\cal M}$ will only include shift counterterms, necessary to realize the Ward identities as the basis of factorization at the local level. 
The jet subdiagrams therefore never include an off-shell loop correction (vertex or self-energy), and no loop polarization counterterms are necessary. 

The explicit factorized one-loop infrared subtraction in Eq.~\eqref{eq:GGsubtraction} is the analog of Eq.~\eqref{eq:EWKsubtraction1} for quark-initiated processes. The one-loop form factor in this case is labeled ``scalar". We discuss the origin of this terminology in Sec.~\ref{sec:scalardecomp} below, but observe that it is independent of the spin states of the incoming gluons. Compared to the case of quark pairs, all dependence on the spin states is contained in infrared finite short distance functions, at least to this loop level. 
In the following, we will construct explicitly the necessary shift counterterms $\Delta {\cal M}$ and the one-loop scalar form factor ${\cal F}_{\rm scalar}$, and confirm that with these definitions all infrared singularities are canceled locally according to Eq.~\eqref{eq:GGsubtraction}. For notational purposes, we combine Eqs.~\eqref{eq:GGsubtraction} and \eqref{eq:bf-M-def} and present our results in the form
\begin{eqnarray}
\label{eq:GGsubtraction-allterms}
{\cal H}_{gg \to {\rm colorless}}^{(2),R}(k,l) &=& {\cal M}_{gg \to {\rm colorless}}^{(2),R}(k,l) - {\cal F}_{\rm scalar}^{(1),R}(k) \,  { {\cal M}}_{gg \to {\rm colorless}}^{(1),R}(l)
\nn\\[2mm]
&\ & - \Delta {\cal M}_{gg \to {\rm colorless}}^{(2),R}(k,l)\, .
\end{eqnarray}
We shall see that the form factor and shift counterterms, separately and in combination, will regularize infrared singularities in separate contributions to the full integrand. 

Having outlined our goal, we are ready to present the particular class of processes with which we will work to illustrate the method, multi-Higgs production. In the Standard Model, a concentration on heavy quark-mediated processes is natural for this set of amplitudes.

\subsection{Amplitudes for multi-Higgs production via gluon fusion}

To simplify the presentation, we will detail concretely the derivation of Eq.~\eqref{eq:GGsubtraction} for processes with multiple Higgs bosons in the final state,
\begin{equation}
\label{eq:multiHiggsprocess}
g(p_1) + g(p_2) \to H(q_1) + \ldots + H(q_n).
\end{equation}
The parentheses denote the momenta of the particles, satisfying momentum conservation
\begin{equation}
p_1 + p_2 = q_1 + \ldots + q_n\,.
\end{equation}

As subgraphs with massless fermion loops are absent in two-loop amplitudes for multi-Higgs production, it is sufficient to work in a theory of QCD with just one heavy quark of mass $m_q$, coupled to an external Higgs field via a Yukawa interaction. We will develop a subtraction formalism in the Feynman gauge, 
a choice that leads to simple integrands. The Lagrangian density is given, including ghost fields $c_a$, by, 
\begin{eqnarray}
\label{eq:Lagrangian}
    {\cal L} &=& -\frac{1}{4} G_{a\mu \nu} G_{a}^{ \mu \nu} -\frac{1}{2} \left(\partial_\mu A_{a}^{\mu} \right)^2  
    +\left(\partial_\mu \bar c_a \right) D_{ab}^\mu c_b 
    \nonumber \\ 
    && 
    +\bar  q \left( 
i \slashed D - m_q
    \right) q - Y_q H \bar q q\, ,
\end{eqnarray}
where $\frac{g_s^2}{4 \pi} = \alpha_s$ is the bare strong coupling constant, $Y_q$ is the bare Yukawa coupling of the massive quark and 
\begin{align}
    G_{a}^{\mu \nu} & = \partial^\mu A^\nu_a - \partial^\nu A^\mu_a + g_s\,f_{abc}A^\mu_b A^\nu_c \,, \nn\\
    D^\mu_{ab} &= \partial^\mu \delta_{ab}- g_s f_{abc} A_c^\mu \,,\nn\\
    D^\mu_{ij} &= \partial^\mu \delta_{ij}- ig_s (T_{c})_{ij} A_c^\mu \,,
\end{align}
the usual definitions for the field strength tensor and covariant derivatives in the adjoint and fundamental representations.
The Lagrangian of Eq.~\eqref{eq:Lagrangian} gives rise to Feynman diagrams where Higgs bosons are emitted directly off the heavy quark to the final state. 
As weak bosons and Higgs self-interactions are omitted in Eq.~\eqref{eq:Lagrangian}, Standard Model diagrams with decays of virtual Higgs and weak gauge bosons decaying to multiple Higgs bosons in the final state are not generated. 
The neglected diagrams can be treated straightforwardly, with the same formalism that we will develop here. For the purpose of demonstrating our method, it will be sufficient to restrict our discussion to the diagrams originating from Eq.~\eqref{eq:Lagrangian} at leading order in the Yukawa coupling $Y_q$ and leading and next-to-leading-order in the QCD coupling $g_s$. 
The perturbative QCD expansion of the amplitudes for multi-Higgs production processes starts from the one-loop order and takes the form, 
\begin{eqnarray}
    M_n &=&  \int \frac{d^D l}{\left(2 \pi \right)^D} {\cal M}_n^{(1)}\left( l \right)
+  \int \frac{d^D l}{\left(2 \pi \right)^D}\frac{d^D k}{\left(2 \pi \right)^D} {\cal M}_n^{(2)}\left( k, l \right) + \ldots
\end{eqnarray}
The subscript $n$ in the amplitude and amplitude integrands indicates the number of Higgs bosons in the final state and the superscript $(m)$, with $m=1,2\,$, indicates the perturbative loop order.
As above, roman $M$ denotes integrated quantities while script, $\cal M$, denotes integrands.  
For simplicity of notation, we absorb factors of $(-iY_q)^n$, which are the same in all diagrams, into the integrands ${\cal M}^{(m)}_n$.

The Feynman diagrams in one-loop amplitudes for gluon-fusion multi-Higgs production are depicted symbolically as
\begin{eqnarray}
\label{eq:BornN}
    {\cal M}_n^{(1)} = 
    \includegraphics[width=0.3\textwidth, page=1,valign=c]{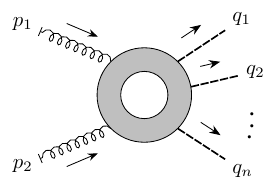}\, ,
\end{eqnarray}
where the initial state gluons and the final state Higgs bosons are attached to a heavy quark loop (gray disk). As an example, the one-loop amplitude for single Higgs production is
\begin{align}
    {\cal M}_1^{(1)} = \includegraphics[width=0.27\textwidth, page=3,valign=c]{figures_standalone/tikz_oneloop.pdf} + \includegraphics[width=0.27\textwidth, page=4,valign=c]{figures_standalone/tikz_oneloop.pdf}.
\end{align}
The Feynman diagrams for the corresponding two-loop amplitudes are depicted in the same fashion as, 
\begin{align}
\label{eq:TwoLoopN}
    {\cal M}_n^{(2)} =& 
    \includegraphics[width=0.27\textwidth, page=23,valign=c]{figures_standalone/tikz_general_diagrams.pdf} + \includegraphics[width=0.27\textwidth, page=3,valign=c]{figures_standalone/tikz_general_diagrams.pdf} + \includegraphics[width=0.27\textwidth, page=4,valign=c]{figures_standalone/tikz_general_diagrams.pdf}\nonumber\\
    &+
    \includegraphics[width=0.255\textwidth, page=5,valign=c]{figures_standalone/tikz_general_diagrams.pdf}
    +\includegraphics[width=0.27\textwidth, page=6,valign=c]{figures_standalone/tikz_general_diagrams.pdf}
    +\includegraphics[width=0.255\textwidth, page=7,valign=c]{figures_standalone/tikz_general_diagrams.pdf}.
\end{align}
The first term on the right-hand side of Eq.~\eqref{eq:TwoLoopN} represents diagrams where a virtual gluon is exchanged between the two incoming, initial state gluons. The second and third terms represent Feynman diagrams where a virtual gluon is exchanged between an initial state gluon and a heavy quark. As examples of diagrams in the first line of Eq.~\eqref{eq:TwoLoopN}, in di-Higgs production we find
\begin{align}
    \includegraphics[width=0.28\textwidth, page=2,valign=c]{figures_standalone/tikz_doubleHiggs.pdf}
    + \includegraphics[width=0.28\textwidth, page=6,valign=c]{figures_standalone/tikz_doubleHiggs.pdf}
    + \includegraphics[width=0.28\textwidth, page=15,valign=c]{figures_standalone/tikz_doubleHiggs.pdf}
    +\ldots\ .
\end{align}
In the second line of Eq.~\eqref{eq:TwoLoopN}, we represent sums of diagrams where a virtual gluon exchange takes place among heavy quarks and diagrams with a quartic gluon vertex. For example, in di-Higgs production we encounter diagrams like
\begin{align}
\includegraphics[width=0.28\textwidth, page=16,valign=c]{figures_standalone/tikz_doubleHiggs.pdf}+ \includegraphics[width=0.28\textwidth, page=17,valign=c]{figures_standalone/tikz_doubleHiggs.pdf} + \includegraphics[width=0.2\textwidth, page=18,valign=c]{figures_standalone/tikz_doubleHiggs.pdf} +\ldots\ .
\end{align}

As indicated in Eq.~\eqref{eq:GGsubtraction-full}, sometimes it will be convenient to factor out the polarization vectors and color of the incoming gluons from the one-loop amplitude. For our one-loop diagrams, the color structure is always the same, $\Tr[T_{a} T_{b}]=1/2\,\delta_{ab}$, where the indices $a,b$ correspond to the color of the incoming gluons $p_1$ and $p_2$. We only factor out the trivial color tensor structure $\delta_{ab}$ from the amplitude and leave the universal factor of $1/2$ in $\widetilde{\cal M}_n^{(1)}$. We write 
\begin{eqnarray}
\label{eq:Mtilde}
    {\cal M}_n^{(1)} \equiv  \, \pol_{1\alpha} \, \pol_{2\beta} \, \delta_{ab}\, {\widetilde {\cal M}}_{n}^{(1) \, \alpha \beta}\, .
\end{eqnarray}
For the full one- and two-loop amplitudes symbolized by ${\cal M}_n^{(m)}$, $m=1,2$, the color is handled implicitly and hence never written out as indices.

The incoming gluons with momenta $p_1$ and $p_2$ have polarization vectors $\pol_1 = \pol(p_1)$ and $\pol_2 = \pol(p_2)$, respectively. The polarisations satisfy the transversality conditions, 
\begin{eqnarray}
\label{eq:polarisations}
\pol_i \cdot p_i = \pol_i \cdot \eta_i =0 \quad {\rm with} \quad p_i \cdot \eta_i \neq 0,   \quad \forall \;  i \in {1,2} \, ,
\end{eqnarray}
where $\eta_i$ are arbitrary light-like reference vectors.

\subsection{Scalar decomposition of the triple-gluon vertex}
\label{sec:scalardecomp}

Diagrams with triple-gluon vertices (the first three terms on the right-hand side of Eq.~\eqref{eq:TwoLoopN}) are the origin of collinear singularities in the two-loop amplitude integrand ${\cal M}_n^{(2)}$. The familiar Feynman rule for the triple-gluon vertex consists of three pairs of terms,
\begin{align}
\label{eq:Vggg}
&\includegraphics[width=0.28\textwidth,page=1,valign=c]{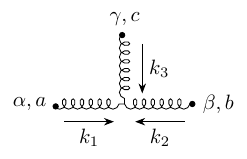}\nn\\
    =& -g_s f_{abc} \left[  \eta^{\alpha\beta} (k_1-k_2)^{\gamma}+  \eta^{\beta\gamma} (k_2-k_3)^{\alpha}+  \eta^{\gamma \alpha} (k_3-k_1)^{\beta} \right] \nn \\[2mm]
    =& -g_s f_{abc} \eta^{\alpha\beta} (k_1-k_2)^{\gamma} -g_s f_{bca}  \eta^{\beta\gamma} (k_2-k_3)^{\alpha} -g_s f_{cab}  \eta^{\gamma \alpha} (k_3-k_1)^{\beta}\, .
\end{align}
In the figure, we indicate the truncation of the incoming lines by dots, a notation that will be used in subsequent sections. In the second relation of Eq.~\eqref{eq:Vggg}, we cyclically permute color indices so that in each term the final color index matches the vector index of the corresponding momentum factor. The distinct terms within the triple-gluon vertex expression of Eq.~\eqref{eq:Vggg} exhibit different behaviors in collinear limits when inserted into diagrams. They also assume different roles in the cancellations among diagrams that ultimately result in the factorization of collinear singularities. Consequently, it is useful to differentiate the terms in Eq.~\eqref{eq:Vggg} and track their contributions to diagrams. We achieve this objective graphically, as follows 
\begin{align}
\label{eq:Vggg_decomposition}
    &\includegraphics[width=0.28\textwidth, page=1,valign=c]{figures_standalone/tikz_scalar_decomp.pdf} \nonumber \\
    &= \includegraphics[width=0.28\textwidth, page=2,valign=c]{figures_standalone/tikz_scalar_decomp.pdf} + \includegraphics[width=0.28\textwidth, page=3,valign=c]{figures_standalone/tikz_scalar_decomp.pdf} + \includegraphics[width=0.28\textwidth, page=4,valign=c]{figures_standalone/tikz_scalar_decomp.pdf},\nn\\
\end{align}
where 
\begin{eqnarray}
\label{eq:Vssg}
\includegraphics[width=0.28\textwidth, page=2,valign=c]{figures_standalone/tikz_scalar_decomp.pdf}
&=& -g_s f_{abc} \eta^{\alpha\beta} (k_1-k_2)^{\gamma} \, . 
\end{eqnarray}
The first color index corresponds to the incoming momentum $k_1$, the second to the subtracted incoming momentum $k_2$ and the last color index matches the vector index of the momenta on the right-hand side. The Feynman rule on the right-hand side of Eq.~\eqref{eq:Vssg} corresponds to the tree-level rule for the interaction of a color-octet scalar and a gluon times the metric, $\eta^{\alpha \beta}$. Motivated by this observation, we will refer to the decomposition of Eq.~\eqref{eq:Vggg_decomposition} as a ``scalar decomposition''. With this graphical decomposition, a diagram with $N_{ggg}$ triple-gluon vertices is represented as the sum of $3^{N_{ggg}}$ diagrams with ``scalar''-``scalar''-gluon vertices. 
For example, we can decompose the following diagram for di-Higgs production with a single triple-gluon vertex into three diagrams with ``scalars'', 
\begin{align}
\label{eq:SDexample1}
     \includegraphics[width=0.3\textwidth, page=4 ,valign=c]{figures_standalone/tikz_doubleHiggs.pdf}   
     = & \includegraphics[width=0.3\textwidth, page=10 ,valign=c]{figures_standalone/tikz_sd_example.pdf} 
    +\includegraphics[width=0.3\textwidth, page=11 ,valign=c]{figures_standalone/tikz_sd_example.pdf} \nn\\
    &+ \includegraphics[width=0.3\textwidth, page=12 ,valign=c]{figures_standalone/tikz_sd_example.pdf}  
\, . 
\end{align}
Similarly, the following diagram with two triple-gluon vertices gives rise to nine diagrams with ``scalars'', 
\begin{align}
\label{eq:SDexample2}
&\includegraphics[width=0.3\textwidth, page=1 ,valign=c]{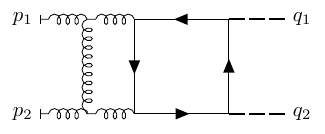} \nn\\
&= 
\includegraphics[width=0.3\textwidth, page=1,valign=c]{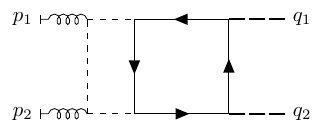}
+\includegraphics[width=0.3\textwidth, page=2,valign=c]{figures_standalone/tikz_sd_example.pdf} 
 + \includegraphics[width=0.3\textwidth, page=3,valign=c]{figures_standalone/tikz_sd_example.pdf}  \nn \\
 &+  \includegraphics[width=0.3\textwidth, page=4,valign=c]{figures_standalone/tikz_sd_example.pdf}
 +\includegraphics[width=0.3\textwidth, page=5,valign=c]{figures_standalone/tikz_sd_example.pdf}  
 + \includegraphics[width=0.3\textwidth, page=6,valign=c]{figures_standalone/tikz_sd_example.pdf} \nn \\
  &+ \includegraphics[width=0.3\textwidth, page=7,valign=c]{figures_standalone/tikz_sd_example.pdf}  
+\includegraphics[width=0.3\textwidth, page=8,valign=c]{figures_standalone/tikz_sd_example.pdf}
 + \includegraphics[width=0.3\textwidth, page=9,valign=c]{figures_standalone/tikz_sd_example.pdf}.\nn\\ 
\end{align}
The dashed lines have the same couplings at vertices as octet scalars. Nevertheless, they retain metric tensors, and it is important to keep in mind that these lines originate from the decomposition of the triple-gluon vertex. Therefore, the dashed lines in propagators should be read with the same rule as of the gluon propagator. We thus define
\begin{align}
\includegraphics[width=0.25\textwidth, page=5,valign=c]{figures_standalone/tikz_scalar_decomp.pdf}  
&=
\includegraphics[width=0.25\textwidth, page=6,valign=c]{figures_standalone/tikz_scalar_decomp.pdf}   
=\includegraphics[width=0.25\textwidth, page=7,valign=c]{figures_standalone/tikz_scalar_decomp.pdf} \nonumber \\  & 
= - i \, \delta_{ab} \, 
    \frac{\eta^{\alpha \beta}}{k^2 + i \delta}
    \, , \quad \delta \to 0^+\, . 
\end{align}
Similarly, dashed incoming legs should be read with the same rule as incoming gluons and be substituted with a polarization vector, 
\begin{equation}  
\includegraphics[width=0.13\textwidth, page=9,valign=c]{figures_standalone/tikz_scalar_decomp.pdf}
\equiv
\includegraphics[width=0.13\textwidth, page=8,valign=c]{figures_standalone/tikz_scalar_decomp.pdf}
= \pol(p)^\alpha \, .    
\label{eq:pol_Feynrules}
\end{equation}
Finally, for a ``scalar''-quark interaction vertex, we  need a Feynman rule that is identical to the quark-gluon interaction vertex, 
\begin{equation} 
\includegraphics[width=0.22\textwidth, page=2,valign=c]{figures_standalone/tikz_Feynrules.pdf} \equiv\includegraphics[width=0.22\textwidth, page=1,valign=c]{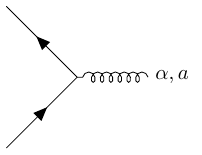}  = -i g_s T_a \gamma^\alpha \, .   
\label{eq:scalar-to-fermion}
\end{equation}

In what follows, we drop the quotation marks for scalars and will refer to the diagrams in Eq.~\eqref{eq:SDexample2} by the sequence of scalar and gluon lines from left to right in the figure. The first diagram on the right-hand side is a ``gluon-gluon" process, the second a ``gluon-scalar" and so on. Finally, we note that these gluon-scalar vertices are different from the vertices of scalar ghosts. At this loop order in the process we study, no ghost propagators are necessary.

\section{Infrared singularities in the two-loop amplitude}
\label{sec:IRsing_channels}

The standard Landau equation, pinch-surface and power-counting analysis of the two-loop diagrams in Eq.~\eqref{eq:TwoLoopN} shows that only two regions can produce infrared singularities \cite{Sterman:1995fz}. These regions occur when an external gluon and its adjacent virtual gluons are collinear. The two collinear regions intersect at a single point in $k$ loop space, which corresponds to soft gluon exchange for diagrams such as in the first term on the right-hand side of Eq.~\eqref{eq:TwoLoopN}.

Applying the scalar decomposition defined in Eq.~\eqref{eq:Vggg_decomposition} to the triple-gluon vertices of the full set of diagrams in Eq.~\eqref{eq:TwoLoopN} facilitates the organization of infrared singularities in the amplitude, making it convenient to match groups of singular terms against appropriate counterterms.
The diagrams involving scalars and gluons are graphically diverse. For example, upon examining the right-hand side of Eq.~\eqref{eq:SDexample1}, we identify two diagrams with a scalar-gluon initial state and one diagram with a gluon-gluon initial state. Similarly, on the right-hand side of Eq.~\eqref{eq:SDexample2}, we observe four diagrams with a scalar-scalar initial state, four diagrams with scalar-gluon or gluon-scalar initial states, and one with a gluon-gluon initial state. Note that scalar lines attach to the fermion loop only through the mediation of the scalar-gluon propagator. The coupling of such lines to the fermions is defined by Eq.~\eqref{eq:scalar-to-fermion}.

The distinction among these diverse ``subprocesses'' (scalar-scalar, scalar-gluon, gluon-scalar, and gluon-gluon) is useful for systematically analyzing the infrared singularities of amplitudes. We will determine that diagrams in the gluon-gluon subprocess do not exhibit soft or collinear singularities. In addition, we will find infrared singularities which can be factored in a select subset of diagrams, which we identify as the scalar form factor subtraction in Eq.~\eqref{eq:GGsubtraction-allterms}. 

\subsection{Analysis by channel}

After applying the decomposition in Eq.~\eqref{eq:Vggg_decomposition} to the diagrams of Eq.~\eqref{eq:TwoLoopN} with triple-gluon vertices, the full two-loop amplitudes can be rewritten as a sum of sixteen distinguishable diagrams. We organize these as 
\begin{eqnarray}
\label{eq:subprocesses}
    {\cal M}^{(2)}_n &=& 
    {\cal M}^{(2)}_{ss \to n}
    +    {\cal M}^{(2)}_{sg \to n}
    +    {\cal M}^{(2)}_{gs \to n}
    +    {\cal M}^{(2)}_{gg \to n},
\end{eqnarray}
where we identify contributions according to the initial states which emerge from Eq.~\eqref{eq:Vggg_decomposition},
\begin{itemize}
    \item scalar-scalar
\begin{eqnarray}
\label{eq:scalar-scalar}
    {\cal M}^{(2)}_{ss \to n} &=&     
        \includegraphics[width=0.27\textwidth, page=8,valign=c]{figures_standalone/tikz_general_diagrams.pdf}
        +\includegraphics[width=0.27\textwidth, page=9,valign=c]{figures_standalone/tikz_general_diagrams.pdf} 
        \nonumber \\ 
        &+&
        \includegraphics[width=0.27\textwidth, page=10,valign=c]{figures_standalone/tikz_general_diagrams.pdf}
        +\includegraphics[width=0.27\textwidth, page=11,valign=c]{figures_standalone/tikz_general_diagrams.pdf}       
\; , 
\end{eqnarray}
\item gluon-scalar
\begin{eqnarray}
\label{eq:gluon-scalar}
    {\cal M}^{(2)}_{gs \to n} &=&  
    \includegraphics[width=0.27\textwidth, page=12,valign=c]{figures_standalone/tikz_general_diagrams.pdf}
    +\includegraphics[width=0.27\textwidth, page=13,valign=c]{figures_standalone/tikz_general_diagrams.pdf}
     \nn\\
    && +\includegraphics[width=0.27\textwidth, page=14,valign=c]{figures_standalone/tikz_general_diagrams.pdf}
    \; , 
\end{eqnarray}
    \item scalar-gluon
\begin{eqnarray}
\label{eq:scalar-gluon}
    {\cal M}^{(2)}_{sg \to n} &=&   
    \includegraphics[width=0.27\textwidth, page=15,valign=c]{figures_standalone/tikz_general_diagrams.pdf}
    +\includegraphics[width=0.27\textwidth, page=16,valign=c]{figures_standalone/tikz_general_diagrams.pdf}\nn\\
    &&+\includegraphics[width=0.27\textwidth, page=17,valign=c]{figures_standalone/tikz_general_diagrams.pdf} 
    \; , 
\end{eqnarray}
    \item and gluon-gluon
\begin{eqnarray}
\label{eq:gluon-gluon}
    {\cal M}^{(2)}_{gg \to n} &=&
   \includegraphics[width=0.27\textwidth, page=18,valign=c]{figures_standalone/tikz_general_diagrams.pdf}
   + \includegraphics[width=0.27\textwidth, page=19,valign=c]{figures_standalone/tikz_general_diagrams.pdf}\nn\\
   && + \includegraphics[width=0.27\textwidth, page=20,valign=c]{figures_standalone/tikz_general_diagrams.pdf}
   +\includegraphics[width=0.25\textwidth, page=5,valign=c]{figures_standalone/tikz_general_diagrams.pdf}\nn\\
    &&+\includegraphics[width=0.27\textwidth, page=6,valign=c]{figures_standalone/tikz_general_diagrams.pdf}
    + \includegraphics[width=0.25\textwidth, page=7,valign=c]{figures_standalone/tikz_general_diagrams.pdf}
   \, ,
\end{eqnarray}
\end{itemize}
where we have included a momentum assignment for gluons and scalars for most of the diagrams.
We now go on to analyze their infrared content.

\subsubsection*{Finiteness of the gluon-gluon channel }

All diagrams in the gluon-gluon channel in Eq.~\eqref{eq:gluon-gluon} are free of collinear and soft singularities. This is due to the following mechanisms:
\begin{itemize}
\item Diagrams with a quartic gluon vertex, two heavy quark fermion loops, or with a virtual gluon exchange in between heavy quarks in a single quark loop do not have all necessary singular propagators in soft and collinear limits to produce a singularity.
\item Diagrams with gluons splitting into scalars vanish due to the transversality of the polarization of the gluons,
Eq.~\eqref{eq:polarisations},
%\[
$\pol_1 \cdot p_1 = \pol_2 \cdot p_2 =0\,.$ 
%\]
\end{itemize}

\subsubsection*{Infrared singularities in the channels with initial state scalars}

All diagrams with one or two scalars in the initial state exhibit infrared singularities locally. However, some of the diagrams, individually or combined with others, have singularities that do not lead to a divergence in the dimensional regulator expansion. These diagrams can be made integrable in $D=4$ dimensions via shift counterterms as in Eq.\ \eqref{eq:bf-M-def}, although not until they are suitably combined, to take advantage of cancellations
between diagrams whose singularities are related by gauge invariance.
We shall refer to these as ``shift-integrable" diagrams and singularities.

\subsection{Shift-integrable infrared singularities}
\label{sec:shift-discussion}

As an example of diagrams in a shift-integrable set, 
consider 
\begin{align}
    {\cal M}^{(2)}_{ss \to gg \to n} &\equiv
    \includegraphics[width=0.27\textwidth, page=8,valign=c]{figures_standalone/tikz_general_diagrams.pdf}\nn\\ 
    &=i\, g_s^2 \, C_A\,\delta_{ab}  \,  \epsilon_1\cdot \epsilon_2\,
    \frac{ (p_1-k)^\alpha (p_2+k)^\beta}{ k^2 \, (k+p_1)^2 \, (k-p_2)^2}\,\widetilde{\mathcal{M}}^{(1)}_{n\,\alpha \beta}\left(l,k+p_1, p_2-k \right) \, , \label{eq:Mssggn}
\end{align}
where ${\widetilde {\cal M}}^{(1)}_{n\,\alpha \beta}\left( l,k+p_1, p_2-k \right)$ has been defined in Eq.~\eqref{eq:Mtilde} as the tensor constructed from the one-loop amplitude by separating its polarization vectors and its (trivial) color tensor. We indicate internal loop momentum and the momenta of the incoming gluons as the arguments of the ${\widetilde {\cal M}}^{(1)}_{n\,\alpha \beta}$ subdiagram. We emphasize that Eq.~\eqref{eq:Mssggn} represents the set of diagrams found by attaching gluons of momentum $k+p_1$ and $p_2-k$ around the fermion loop in all possible ways. 
This diagram is singular in both collinear limits, 
$k=-xp_1$ and $k=xp_2$ with $0< x<1$, as well as in the soft, $k\to 0$ limit. In each collinear region, one of the incoming lines splits into two collinear lines.

Explicitly, we can see that in a collinear limit, for example as $k$ becomes parallel to $p_1$, the divergent part of the diagrams is given by, 
\begin{align}
\label{eq:exampleSingInteg}
     &\lim_{k = -x p_1}  {\cal M}^{(2)}_{ss \to gg \to n}\nn\\ &= i\, g_s^2 \, C_A\,\delta_{ab}  \, \epsilon_1\cdot \epsilon_2\,
    \frac{ (p_2+k)^\beta}{ k^2 \, (k+p_1)^2 \, (k-p_2)^2}\, \frac{1+x}{1-x}\,(k+p_1)^\alpha \,\widetilde{\mathcal{M}}^{(1)}_{n\,\alpha \beta}\left(l,k+p_1, p_2-k \right) \, ,
\end{align}
where we have used
\begin{align}
    \lim_{k = -x p_1} (p_1-k)^\alpha = \lim_{k = -x p_1} (1+x)\,p_1^\alpha = \lim_{k = -x p_1} \frac{1+x}{1-x} (k+p_1)^\alpha\,.
\end{align}
Because of gauge invariance, the singularity at $k=-xp_1$ in Eq.~\eqref{eq:exampleSingInteg} is shift-integrable. 
Note that in this collinear limit, the gluon with momentum $k+p_1$ entering the fermion-loop subgraph acquires a longitudinal polarization vector (due to the momentum vector $(k+p_1)^\alpha$ in the numerator of Eq.~\eqref{eq:exampleSingInteg}). As we will review in Sec.~\ref{sec:one-loop_WI}, the one-loop amplitude obeys the familiar QED Ward identities,
\begin{align}
\label{eq:WIloopAsec3}
    &\int d^D l\; (k+p_1)^{\alpha} \, {\widetilde {\cal M}}^{(1)}_{n\,\alpha\beta}\left(l,k+p_1, p_2-k\right) =0\, , \nn\\
& \int \mathrm{d}^D l\; (p_2-k)^{\beta} \,  {\widetilde {\cal M}}^{(1)}_{n\,\alpha\beta}\left( l,k+p_1, p_2-k\right) = 0 \, ,
\end{align}
which hold after all the diagrams in the set are summed over, and the $l$ integration is carried out. It is then evident that after we carry out the integration over the fermion loop, the collinear and soft singularities in the $k$ integration cancel among the set of diagrams that define ${\cal M}^{(2)}_{ss\to gg \to n}$. 
The identities, Eq.\ \eqref{eq:WIloopAsec3} hold in this QCD calculation because all of these diagrams have the same color factor, and so are in one-to-one correspondence with QED diagrams.
Thus in this case, as in QED, a scalar-polarized vector decouples from a fermion loop for arbitrary polarizations of additional vector lines. This is not generally the case for nonabelian fermion loops. 

With a similar analysis, we will find below that in the scalar-scalar, gluon-scalar and scalar-gluon channels, the infrared singularities of other combinations of diagrams are shift-integrable. 

\subsection{Synopsis of singular diagrams}

After the scalar decomposition of Eq.~\eqref{eq:Vggg_decomposition}, we can separate the two-loop amplitude into infrared singular and infrared finite diagrams. In fact, a further subdivision within the category of infrared singular diagrams emerges from our previous analysis of collinear limits.
 We can cast 
the amplitude as the sum of three contributions,
\begin{eqnarray}
\label{eq:Mir+Mrest}
    {\cal M}_{n}^{(2)} ={\cal M}_{n,{\rm IR}}^{(2)\text{,fact}} + 
    {\cal M}_{n,{\rm IR}}^{(2)\text{,shift}}
    + {\cal M}_{n,{\rm \overline{IR}}}^{(2)} \, ,
\end{eqnarray}
where the subscripts IR identify the two terms with IR singularities, and $\overline{\rm IR}$ a term that is locally integrable in the infrared. We defer the treatment of ultraviolet regularization to Sec.~\ref{sec:UVCT}, which does not change the construction that follows.

The first term, ${\cal M}_{n,{\rm IR}}^{(2)\text{,fact}}$, in Eq.~\eqref{eq:Mir+Mrest} consists of three diagrams from Eq.~\eqref{eq:scalar-scalar}, whose collinear singularities are related by gauge invariance.
\begin{align}
\label{eq:Mn2IRfact}
{\cal M}_{n,{\rm IR}}^{(2)\text{,fact}}  &=
 {\cal M}^{(2)}_{ss \to ss \to n}
 +  {\cal M}^{(2)}_{sg \to (sg)g \to n}
 +  {\cal M}^{(2)}_{gs \to g(sg) \to n} \nn\\[4mm]
& =
\includegraphics[width=0.25\textwidth, page=9,valign=c]{figures_standalone/tikz_general_diagrams.pdf}   
+\includegraphics[width=0.25\textwidth, page=15,valign=c]{figures_standalone/tikz_general_diagrams.pdf}
+\includegraphics[width=0.25\textwidth, page=12,valign=c]{figures_standalone/tikz_general_diagrams.pdf} 
 \, .  
\end{align}
We will show, in Section~\ref{sec:factorization}, that their collinear singularities factorize and that all their infrared singularities are approximated by the form factor counterterm (second term on the right-hand side) of Eq.~\eqref{eq:GGsubtraction} and an appropriate shift counterterm.

The second term in Eq.~\eqref{eq:Mir+Mrest},
${\cal M}_{n,{\rm IR}}^{(2)\text{,shift}}$, is given by the remaining seven diagrams in Eqs.~\eqref{eq:scalar-scalar}, \eqref{eq:gluon-scalar} and \eqref{eq:scalar-gluon},
\begin{align}
{\cal M}_{n,{\rm IR}}^{(2)\text{,shift}} =& \includegraphics[width=0.25\textwidth, page=8,valign=c]{figures_standalone/tikz_general_diagrams.pdf} +   \includegraphics[width=0.25\textwidth, page=10,valign=c]{figures_standalone/tikz_general_diagrams.pdf} +  \includegraphics[width=0.25\textwidth, page=11,valign=c]{figures_standalone/tikz_general_diagrams.pdf}\nonumber \\
 +&\includegraphics[width=0.25\textwidth, page=16,valign=c]{figures_standalone/tikz_general_diagrams.pdf} +  \includegraphics[width=0.25\textwidth, page=17,valign=c]{figures_standalone/tikz_general_diagrams.pdf}  + \includegraphics[width=0.25\textwidth, page=13,valign=c]{figures_standalone/tikz_general_diagrams.pdf} \nonumber\\
+&\includegraphics[width=0.25\textwidth, page=14,valign=c]{figures_standalone/tikz_general_diagrams.pdf}\, .
\label{eq:ssfusdiag_generic_all_diags-0}
\end{align}
This set of diagrams will be treated in Sec.~\ref{sec:NonLocalSings}. It includes all diagrams that exhibit soft and collinear singularities within the integrand but are, in fact, infrared finite after integration due to the Ward identity in Eq.~\eqref{eq:WIloopAsec3}.
Despite the infrared finiteness of the integral, it is necessary to eliminate these local singularities for numerical integration purposes through a shift subtraction. In Sec.~\ref{sec:NonLocalSings}, we will demonstrate that in singular collinear limits, after all cancellations among diagrams are accounted for, there remain singularities in the form of a difference of integrands with a relative shift in their loop momenta. These non-local but shift-integrable singularities can be eliminated by making use of the Ward identities and introducing shift counterterms, as in Eqs.~\eqref{eq:bf-M-def} and \eqref{eq:bf-Delta-M-vanishes}, akin those in Refs.~\cite{Anastasiou:2020sdt,Anastasiou:2022eym}.

The final, locally integrable term, $ {\cal M}_{n,{\rm \overline{IR}}}^{(2)}$ in Eq.~\eqref{eq:Mir+Mrest} is specified by all the diagrams of Eq.~\eqref{eq:gluon-gluon}, and needs no further treatment for local integrability.

We will show that each term in Eq.~\eqref{eq:Mir+Mrest} gives a well-defined contribution to the locally-integrable hard scattering function ${\cal H}$. Following the notation of Eq.~\eqref{eq:GGsubtraction-allterms}, the three terms each contribute to the hard scattering separately, 
\begin{align}
\label{eq:calH-threeterms}
    {\cal H}_{n}^{(2)}(k,l)
    = 
    {\cal H}_{{n, {\rm IR}}}^{(2),{\rm fact}}(k,l)
    +
    {\cal H}_{{n, {\rm IR}}}^{(2),{\rm shift}}(k,l)
    +
    {\cal H}_{{n, {\rm \overline{IR}}}}^{(2)}(k,l)\, .
\end{align}
The terms on the right of Eq.~\eqref{eq:Mir+Mrest}, and correspondingly their locally infrared finite contributions in Eq.~\eqref{eq:calH-threeterms}, are each specified by the diagrams of Eqs.~\eqref{eq:scalar-scalar} - \eqref{eq:gluon-gluon}. As we shall see, the first requires both a form factor counterterm and a shift counterterm, while the second requires only a shift counterterm. The third is IR finite on its own, without a subtraction,
\begin{align}
\label{eq:GGsubtraction-termbyterm}
{\cal H}_{{n, {\rm IR}}}^{(2),{\rm fact}}(k,l) &= 
{\cal M}_{n,{\rm IR}}^{(2),\text{fact}}(k,l) - {\cal F}_{\rm scalar}^{(1)}(k) \,  { {\cal M}}_{n}^{(1)}(l)
- \Delta {\cal M}_{n,{\rm IR}}^{(2),\text{fact}}(k,l)
\, ,
\nn\\[2mm]
{\cal H}_{n,{\rm IR}}^{(2),\text{shift}}(k,l) &= 
{\cal M}_{n,{\rm IR}}^{(2),\text{shift}}(k,l) 
- \Delta {\cal M}_{n,{\rm IR}}^{(2),\text{shift}}(k,l)\, , 
\nn\\[2mm]
{\cal H}_{n,{\rm\overline{IR}}}^{(2)}(k,l) 
&= 
{\cal M}_{n,{\rm \overline{IR}}}^{(2)} (k,l)
\, .
\end{align}
In the following two sections we will see how most of the singular contributions in these sums cancel among the unmodified diagrams, leaving terms that either factorize as shown, or can be canceled by shift counterterms.
Finally, as noted above, we will describe the ultraviolet subtractions necessary to make these functions integrable at infinite loop momenta in Sec.~\ref{sec:UVCT}. They will be applied to all the terms in ${\cal H}^{(2)}$, including the infrared subtractions identified here.

\section{Factorizable collinear singularities}
\label{sec:factorization}

In this section, we will demonstrate that all collinear and soft singularities in the infrared singular diagrams
${\cal M}_{n,{\rm IR}}^{(2)\text{,fact}}$ of Eq.~\eqref{eq:Mn2IRfact} can be factorized locally after summing all diagrams. We will find that subtracting a scalar form factor times the one-loop amplitude averaged over two momentum flows will provide a contribution to the locally-integrable short-distance function ${\cal H}^{(2)}_{gg\to {\rm colorless}}$. The subtraction can be expressed as the combination of a form factor times the one-loop amplitude and a shift counterterm. 

To study the factorizable diagrams of ${\cal M}_{n,{\rm IR}}^{(2)\text{,fact}}$, we will assign suitable flows for the loop momenta. Subsequently, we will revisit the essential Ward identities governing the behavior of longitudinally polarized gluons. Using these identities, we will then establish the factorization of collinear singularities. The soft singularities will be automatically accounted for in the same procedure.

\subsection{Assignment of momentum flows}

In loop amplitudes, loop momenta can be freely subjected to shifts in the presence of a dimensional regulator. At collinear limits, these shifts induce alterations in the pattern of cancellations within the sum of diagrams. Given our objective to subtract or cancel collinear singularities prior to integration in exactly $D=4$ dimensions, the assignment of loop momentum flows must satisfy additional constraints. 

We will assign loop momentum flows in the diagrams in ${\cal M}_{n,{\rm IR}}^{(2)\text{,fact}}(k,l)$, as defined in Eq.~\eqref{eq:Mn2IRfact}, in a manner that permits local cancellations in collinear limits at each point defined by the integration variables $k,l$ in the integration domain. Our assignment of loop momentum labels $k,l$ in the two-loop diagrams will be coordinated with the assignment of the loop momentum $l$ made for the corresponding diagrams in the Born amplitude ${\cal M}_{n}^{(1)}(l)$.

\subsubsection*{Momentum flows in the one-loop amplitude}

The diagrams that define the Born amplitude can be represented as 
\begin{align}
    \mathcal{M}_n^{(1)}(l,p_1,p_2) \equiv \sum_{\text{perms } \{q_1,\ldots,q_n\}} \sum_{r=0}^n
\includegraphics[width=0.35\textwidth, page=1,valign=c]{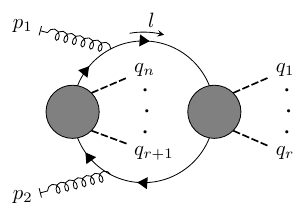}\, .
   \label{eq:oneloopdiag_qlabels}
\end{align}
Here, the blob where two internal fermion lines and several external Higgs bosons attach represents the ordered emission of scalars in Eq.~\eqref{eq:oneloopdiag_qlabels},
\begin{align}
    \includegraphics[width =0.15\textwidth, page=2,valign=c]{figures_standalone/tikz_diags.pdf} 
    = \begin{cases}
        \includegraphics[width =0.3\textwidth, page=3,valign=c]{figures_standalone/tikz_diags.pdf} & i<j,\\
        \includegraphics[width =0.11\textwidth, page=5,valign=c]{figures_standalone/tikz_diags.pdf} & i=j.
    \end{cases}
\end{align}
The one-loop amplitude is given by the sum over $r$, where $r$ splits the Higgs boson external legs into two groups, as well as a sum over all permutations of the set of external leg momenta $\{q_1,\ldots ,q_n\}$.
We assign a loop momentum label $l$ to the heavy quark propagator leaving the vertex where $p_1$ attaches to the loop, in the direction of the charge flow. Finally, we note that ${\cal M}_n^{(1)}$ is defined to include polarizations for the incoming gluons, $p_1$ and $p_2$.

\subsubsection*{Momentum flows in the two-loop diagrams} 
\label{sec:momflowsM2fact}

As we shall see, all terms of ${\cal M}_{n,{\rm IR}}^{(2)\text{,fact}}$, the right-hand side of Eq.~\eqref{eq:Mn2IRfact}, have collinear singularities, which partially cancel in their sum, with a remainder that is either factorized from the quark loop or which can be removed by a shift counterterm. All these diagrams feature a virtual gluon that connects to one or both external legs, of momentum $p_1$ or $p_2$. To ensure localized collinear cancellations, we assign, in all diagrams, a momentum $k$ to this virtual gluon, directed towards external leg $p_1$ or/and away from external leg $p_2$, as already introduced in Eq.~\eqref{eq:Mn2IRfact}. When the momentum $k$ enters the quark loop subgraph, its direction aligns with the charge flow arrow.

For the second loop momentum $l$, flowing exclusively within the fermion loop, we make an assignment analogous to that in the one-loop amplitude. Specifically, upon removing the $k$ gluon from the two-loop diagrams, the momentum flow aligns with that of a corresponding one-loop diagram in Eq.~\eqref{eq:oneloopdiag_qlabels}. The graphical equations below illustrate these momentum flows for the three classes of diagrams.
\begin{align}
\label{eq:momentumflow_A}
{\cal M}^{(2)}_{ss \to ss \to n}
& = \includegraphics[width=0.38\textwidth, page=7,valign=c]{figures_standalone/tikz_diags.pdf},
\end{align}
\begin{align}
\label{eq:momentumflow_B}
  {\cal M}^{(2)}_{sg \to  (sg)g \to n}
& =  \includegraphics[width=0.38\textwidth, page=9,valign=c]{figures_standalone/tikz_diags.pdf} + \includegraphics[width=0.38\textwidth, page=8,valign=c]{figures_standalone/tikz_diags.pdf},
\end{align}
\begin{align}
\label{eq:momentumflow_C}
 {\cal M}^{(2)}_{gs \to g(sg) \to n}   &= \includegraphics[width=0.38\textwidth, page=11,valign=c]{figures_standalone/tikz_diags.pdf} + \includegraphics[width=0.38\textwidth, page=10,valign=c]{figures_standalone/tikz_diags.pdf}.
\end{align}
In the above, we implicitly sum over all permutations of the external leg momenta, and an index $r$, which partitions the electroweak vertices relative to attachments of the gluon and scalar lines that carry the external polarizations, analogously to the one-loop amplitude in Eq.~\eqref{eq:oneloopdiag_qlabels}. 

\subsection{Ward identities}

The factorization of collinear singularities will follow from gauge symmetry, by exploiting the corresponding Ward identities. Let us consider explicitly the limit $k \parallel p_1$, in which the diagrams ${\cal M}^{(2)}_{ss \to ss \to n}$ in Eq.~\eqref{eq:momentumflow_A} and ${\cal M}^{(2)}_{sg \to (sg)g \to n}$ in Eq.~\eqref{eq:momentumflow_B} are singular. Each of these diagrams has a scalar-scalar-gluon vertex proportional to a factor $(2p_1 + k)^{\alpha}$. 
This vector contracts either with an analogous factor from a second scalar-scalar-gluon vertex in ${\cal M}^{(2)}_{ss \to ss \to n}$ or with a gamma matrix from a quark-quark-gluon vertex in the heavy quark loop in ${\cal M}^{(2)}_{sg \to (sg)g \to n}$. The sum of these singular graphs thus takes the form, 
\begin{align}
{\cal M}^{(2)}_{ss \to ss \to n} + 
{\cal M}^{(2)}_{sg \to (sg)g \to n }
&= g_sf_{acd}\frac{(2 p_1 + k)_\delta}{k^2 (k+p_1)^2} 
    \, {\cal V}_{cd}^\delta (k, l) 
    \nn\\
    &
    \equiv 
    \includegraphics[width=0.3\textwidth, page=27,valign=c]{figures_standalone/tikz_diags.pdf},
    \label{eq:first-Valpha}
\end{align}
where the prefactor in the first relation is found from the conventions for the scalar-scalar-gluon vertex in Eq.~\eqref{eq:Vssg}, and to be specific, includes $(-i)^2$ from the exhibited propagators. We collect the remaining factors in ${\cal V}^\delta_{cd} (k, l) $.
In the collinear limit, $k =-x p_1$, we observe that the vector contracted with ${\cal V}^\delta$ can be reexpressed as
\begin{align}
    (2p_1+k)_\delta \to \frac{(2-x)}{x}(-k)_\delta = \frac{(2p_1+ k)\cdot \chi}{(-k)\cdot \chi} (-k)_\delta,
\end{align}
where $\chi$ is some arbitrary reference vector that fulfills $p_1\cdot \chi \neq 0$. 
We find that the collinear-singular integral occurs when the gluon carries an exactly longitudinal polarization into the sum of diagrams, ${\cal V}^\delta$.
We exhibit this graphically as
\begin{eqnarray}
    \includegraphics[width=0.3\textwidth, page=27,valign=c]{figures_standalone/tikz_diags.pdf}
    &\longrightarrow& 
    g_sf_{acd}
    \frac{(2p_1+k)\cdot \chi }{(-k)\cdot\chi\,  k^2 \left( k+p_1\right)^2 } \, 
    \left[  
    - k_\delta {\cal V}_{cd}^\delta \left( k, l\right)
    \right]_{k \parallel p_1}
    \nonumber \\
    &\equiv& 
    \includegraphics[width=0.3\textwidth, page=28,valign=c]{figures_standalone/tikz_diags.pdf}
    \label{eq:collvertexrule}\, .
\end{eqnarray}
Here the dot joining the gluon coil with the dashed line ending in an arrow represents the following ``collinear approximations" on the polarization tensor of the propagator of the collinear gluon $k$ in the two collinear limits,
\begin{alignat}{2}
    &\frac{-i\, \delta_{md}}{k^2} \eta^{\mu\delta}\longrightarrow \,\frac{-i \,\delta_{md}}{k^2}\frac{\chi^\mu \,k^\delta}{k\cdot \chi} \, \equiv \includegraphics[width=0.32\textwidth, page=21,valign=c]{figures_standalone/tikz_Ward_new.pdf} \,.
    \label{eq:coll_approx}
\end{alignat}
The index $\delta$ on $k^\delta$, corresponding to the dot-dashed-arrow in the figure on the right-hand side of Eq.\ \eqref{eq:collvertexrule}, is summed against a vertex in ${\cal V}^\delta$. The index $\mu$ is summed against the vertex next to the incoming line, to which the gluon becomes collinear. In the above equation, we observe a singularity arising from the denominators of the scalar and gluon lines. However, it is also evident that the virtual gluon acquires a longitudinal polarization $-k^{\delta}$ that prompts a treatment via Ward identities.

For our derivation, the necessary Ward identities concern three-point correlation functions at tree level. For the quark-quark-gluon, we have 
\begin{eqnarray}
\label{eq:qqgWIexplicit}
    k^\alpha \, 
    \frac {i}{\slashed l+\slashed k} \, 
    \left( -i g_s\, T_c\,  \gamma_\alpha  
    \right)\, 
    \frac {i}{\slashed l}
    =   \left( g_s\, T_c \right)\, \frac{i}{\slashed l} + \left(-g_s\, T_c \right) \, \frac{i}{\slashed l +\slashed k}\,  ,
    \label{eq:Wardid1}
\end{eqnarray}
which we represent graphically~\footnote{Similar graphical notations are commonly used in the literature for analyzing Ward identities; see, for example, Refs. \cite{tHooft:1971akt,Sterman:1993hfp,Peskin:1995ev,Schwartz:2014sze}.}
\begin{align}
    \includegraphics[width=0.24\textwidth, page=1,valign=c]{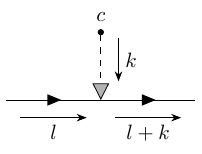} =\raisebox{0.2mm}{\includegraphics[width=0.24\textwidth, page=2,valign=c]{figures_standalone/tikz_Ward_new.pdf}} \hspace{-1cm} + \hspace{-1cm}\includegraphics[width=0.24\textwidth, page=3,valign=c]{figures_standalone/tikz_Ward_new.pdf} . 
    \label{eq:Wardid_qqg_eq}
\end{align}
The dot-dashed line with an arrow on the left-hand side represents one part of the collinear approximation in Eq.~\eqref{eq:coll_approx}. It is a truncated external or internal gluon (indicated by the dot) with a longitudinal, or equivalently scalar, polarization $\pol^\alpha = k^\alpha$ (indicated by the arrow). This polarization is, of course, unphysical, but as seen above, it plays an important role in the occurrence of infrared singularities. 
The vertices on the right-hand side of Eq.~\eqref{eq:Wardid_qqg_eq} are defined as 
\begin{align}
\label{eq:gqqvertexWard}
\includegraphics[width=0.072\textwidth, page=4,valign=c]{figures_standalone/tikz_Ward_new.pdf} \equiv g_s T_c \, , \qquad \includegraphics[width=0.072\textwidth, page=5,valign=c]{figures_standalone/tikz_Ward_new.pdf} \equiv - g_s T_c \hfill\, .
\end{align}
A cross on the quark leg symbolizes the cancellation of a propagator, as in the terms on the right-hand side of Eq.~\eqref{eq:qqgWIexplicit}. Accounting for the relative minus sign of the two terms, a cross on a quark line followed by an outgoing charge arrow at the vertex indicates an extra multiplication by (-1). The double lines represent the insertion of the momentum $k$ and specify the presence of the color generator to which the gluon's color attaches. 

For the scalar-scalar-gluon vertex, the Ward identity reads 
\begin{align}
k^\rho &\,  
 \frac{-i \eta^{\alpha \mu}\, \delta_{am}}{l^2} \, 
\left[ -g_s \, \eta_{\mu \nu} \, f_{mnc} \left(2 l+k \right)_{\rho} \right] \, 
     \frac{-i \eta^{\nu \beta}\,\delta_{nb}}{(l+k)^2}
    \nonumber \\ 
  &=  %\hspace{1cm}
      \left(i g_s  f_{mbc}\,\eta_{\mu\beta}\right)\frac{- i \eta^{\alpha \mu}\,\delta_{am}}{l^2} + \left( i g_s  f_{nac} \,\eta_{\alpha\nu} \right) \frac{- i \eta^{\nu \beta}\,\delta_{nb}}{(l+k)^2} 
     \,.
\end{align}
Graphically, we represent the above identity as
\begin{align}
    \includegraphics[width=0.34\textwidth, page=6,valign=c]{figures_standalone/tikz_Ward_new.pdf}  =  \raisebox{0.2mm}{   \includegraphics[width=0.29\textwidth, page=17,valign=c]{figures_standalone/tikz_Ward_new.pdf} } \hspace{-0.2cm}+ \hspace{-0.2cm}    \includegraphics[width=0.29\textwidth, page=18,valign=c]{figures_standalone/tikz_Ward_new.pdf}. 
    \label{eq:Wardid_ssg_eq}
\end{align}
In the above, we have introduced the vertex 
\begin{align}
\label{eq:Wardid_ssg}
\includegraphics[width=0.18\textwidth, page=19,valign=c]{figures_standalone/tikz_Ward_new.pdf} = i g_s f_{abc}\,\eta_{\alpha \beta}\,,
\end{align}
where the color indices are always in the order: uncanceled line, followed by canceled line, followed by the incoming double lined particle. 
Note that, if a longitudinal gluon attaches to an external scalar leg ($p^2=0$), 
Eq.~\eqref{eq:Wardid_ssg_eq} reduces to 
\begin{eqnarray}
    \includegraphics[width=0.34\textwidth, page=15,valign=c]{figures_standalone/tikz_Ward_new.pdf} = \raisebox{0.2mm}{\includegraphics[width=0.23\textwidth, page=20,valign=c]{figures_standalone/tikz_Ward_new.pdf} }\hspace{-0.5cm} . 
\end{eqnarray}

It is easy to verify an identity due to color conservation, 
\begin{align}
    \label{eq:colorconservationA}
    \includegraphics[width=0.17\textwidth, page=10,valign=c]{figures_standalone/tikz_Ward_new.pdf} + \includegraphics[width=0.17\textwidth, page=11,valign=c]{figures_standalone/tikz_Ward_new.pdf} = 0\, ,
\end{align}
where in the above, the line with broad dashes represents a Higgs boson, emitted from a heavy quark line. Similarly, 
\begin{align}
    \label{eq:colorconservationB}
    \includegraphics[width=0.17\textwidth, page=12,valign=c]{figures_standalone/tikz_Ward_new.pdf} + \includegraphics[width=0.17\textwidth, page=13,valign=c]{figures_standalone/tikz_Ward_new.pdf} + \includegraphics[width=0.17\textwidth, page=14,valign=c]{figures_standalone/tikz_Ward_new.pdf} = 0\, ,
\end{align}
where the short dashed line represents a massless scalar, originating from the scalar decomposition of a triple-gluon vertex. 

Inserting a longitudinal gluon at all positions in a quark line from which a number of Higgs bosons (or, more generally, other color-neutral particles) are emitted results in cancellations among diagrams, 
\begin{align}
\label{eq:scalargluononfermionline}
& \includegraphics[width=0.35\textwidth, page=2,valign=c]{figures_standalone/tikz_line.pdf} 
+\includegraphics[width=0.35\textwidth, page=3,valign=c]{figures_standalone/tikz_line.pdf} \nn\\
& 
+ \ldots
+\includegraphics[width=0.35\textwidth, page=4,valign=c]{figures_standalone/tikz_line.pdf} 
+\includegraphics[width=0.35\textwidth, page=5,valign=c]{figures_standalone/tikz_line.pdf} \nonumber\\
 & = \includegraphics[width=0.3\textwidth, page=6,valign=c]{figures_standalone/tikz_line.pdf} + \includegraphics[width=0.3\textwidth, page=7,valign=c]{figures_standalone/tikz_line.pdf}. 
\end{align}
Note that the local cancellation only occurs if the momentum assignment of the quark lines match in each term. Graphically, we represent the above well known tree-level 
Ward identity as
\begin{eqnarray}
\label{eq:WI-blob}
\includegraphics[width=0.15\textwidth, page=1,valign=c]{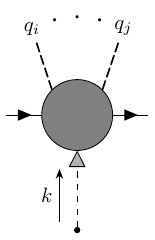} = 
    \raisebox{3.5mm}{\includegraphics[width=0.17\textwidth, page=8,valign=c]{figures_standalone/tikz_line.pdf} + \includegraphics[width=0.17\textwidth, page=9,valign=c]{figures_standalone/tikz_line.pdf}} .
\end{eqnarray}

\subsection{Collinear limits and factorizable singularities}

We will now study the collinear singularities of the three types of diagrams of ${\cal M}_{n,{\rm IR}}^{(2)\text{,fact}}$, for which we have assigned the momentum flows of Eqs.~\eqref{eq:momentumflow_A}, \eqref{eq:momentumflow_B} and \eqref{eq:momentumflow_C}. 
We will demonstrate that in the collinear limits, $k \parallel p_1$ and $k \parallel p_2$, the singularities factorize locally for our choice of momentum flows. In the collinear approximation, the longitudinally polarized gluon is graphically denoted by the gluon-dot-dashed line with a gray arrow as defined in Eq.~\eqref{eq:coll_approx}.
Explicitly, the contributing diagrams in the $k\parallel p_1$ limit are, 
\begin{align}
    &\includegraphics[width=0.3\textwidth, page=31,valign=c]{figures_standalone/tikz_diags.pdf}
 \nonumber \\ 
 & \hspace{1cm}=   
 \includegraphics[width=0.4\textwidth, page=19,valign=c]{figures_standalone/tikz_diags.pdf}
+ \includegraphics[width=0.4\textwidth, 
page=21,valign=c]{figures_standalone/tikz_diags.pdf} \nn\\
&\hspace{1cm}
+ \includegraphics[width=0.4\textwidth, page=20,valign=c]{figures_standalone/tikz_diags.pdf}.
\end{align}

Utilizing the Ward identities from Eqs.~\eqref{eq:Wardid_ssg_eq} and \eqref{eq:scalargluononfermionline} (equivalently, Eq.~\eqref{eq:WI-blob}), it becomes straightforward to demonstrate the factorization of the collinear singularity in the limit $k \parallel p_1$. We write
\begin{align}
&\includegraphics[width=0.3\textwidth, page=31,valign=c]{figures_standalone/tikz_diags.pdf}
\nonumber \\ 
&\hspace{1cm}= \includegraphics[width=0.4\textwidth, page=22,valign=c]{figures_standalone/tikz_diags.pdf} \nn \\
&\hspace{1cm} + \left(
 \includegraphics[width=0.4\textwidth, page=25,valign=c]{figures_standalone/tikz_diags.pdf}
+ \includegraphics[width=0.4\textwidth, page=26,valign=c]{figures_standalone/tikz_diags.pdf}
\right)
\nn \\
&\hspace{1cm}+ \left(
\includegraphics[width=0.4\textwidth, page=23,valign=c]{figures_standalone/tikz_diags.pdf}
+ \includegraphics[width=0.4\textwidth, page=24,valign=c]{figures_standalone/tikz_diags.pdf}
\right). \nn \\
\end{align}
The first, the third and the last diagram on the right-hand side of the above equation cancel by the identity in Eq.~\eqref{eq:colorconservationB}. 
The remaining terms, 
\begin{align}
&\includegraphics[width=0.3\textwidth, page=31,valign=c]{figures_standalone/tikz_diags.pdf}
 \nonumber \\ 
 &=
\includegraphics[width=0.4\textwidth, page=30,valign=c]{figures_standalone/tikz_diags.pdf} \quad
 + \includegraphics[width=0.4\textwidth, page=29,valign=c]{figures_standalone/tikz_diags.pdf}
\label{eq:genampinkp1limit_pic}
\end{align}
have a collinear singularity that factorizes locally in terms of the one-loop amplitudes and external leg corrections, up to a shift in loop momentum $l$ in the first term on the right.

Regarding overall constant factors, the Ward identity notation in Eqs.~\eqref{eq:Wardid_qqg_eq} and \eqref{eq:gqqvertexWard} leads to a minus sign in the second diagram due to outgoing charge arrow from the vertex with a cross. Additionally the trace of the color generators differs by a swap of two generators in the diagrams. Including the constant prefactor in Eq.~\eqref{eq:first-Valpha}, we find the overall factors of the two diagrams from
\begin{align}
    g_sf_{acd}\,g_s\,\Tr(T_b T_c T_d) = g_sf_{acd}\,(-g_s)\,\Tr(T_b T_d T_c) = \frac{i\, g_s^2}{4}\, C_A \, \delta_{ab},
\end{align}
where index $b$ belongs to the color of the incoming gluon with momentum $p_2$. Thus, the two diagrams on the right of Eq.~\eqref{eq:genampinkp1limit_pic} give identical constant factors.

The momentum factors in the integrand, as given in Eq.~\eqref{eq:first-Valpha} and \eqref{eq:collvertexrule}, are also the same for both diagrams, which we can rewrite when $k =-xp_1$, taking $\chi=p_2$, as
\begin{eqnarray}
   \frac{(2p_1+k)\cdot \chi }{(-k)\cdot\chi\,  k^2 \left( k+p_1\right)^2 }
   \rightarrow -
   \frac{1}{k^2(k+p_1)^2}\frac{\left(2p_1+k \right )\cdot p_2}{p_2\cdot k}\, .
\end{eqnarray}
The remaining momentum integrand associated with the fermion loop in Eq.~\eqref{eq:genampinkp1limit_pic} is identical for each diagram to the corresponding contribution to the one-loop amplitude, Eq.~\eqref{eq:oneloopdiag_qlabels}.
Absorbing the color factor $\frac{1}{2}\delta_{ab}$ into the fermion loop momentum trace, we then find for the collinear approximation in Eq.~\eqref{eq:genampinkp1limit_pic} 
\begin{align}
    \lim_{k \parallel p_1} 
   {\cal M}_{n,{\rm IR}}^{(2)\text{,fact}} 
  =&-i\, g_s^2 \,\frac{C_A}{2}\, \frac{1}{k^2(k+p_1)^2}   \frac{(k+2p_1)\cdot p_2}{k\cdot p_2} \nonumber\\
 & \qquad \times
 \left( {\mathcal{M}}_n^{(1)}(l,p_1,p_2) + {\mathcal{M}}_n^{(1)}(l+k,p_1,p_2)\right)\, , 
 \label{eq:genampinkp1limit}
\end{align}
with the full one-loop amplitude ${\cal M}_n^{(1)}$ defined by Eq.~\eqref{eq:oneloopdiag_qlabels}.
Note that the non-singular (hard function) factor contains the one-loop amplitude averaged over two momentum flows, shifted by $l\to l+k$ with respect to each other. We have exhibited the dependence of the incoming momenta, $p_1$ and $p_2$, to emphasize that dependence on $k$ is only through a shift in the loop momentum $l$ and that all external momenta in ${\cal M}_n^{(1)}$ are the same in both terms.

Because of the factorization property in Eq.~\eqref{eq:genampinkp1limit}, a single infrared counterterm will cancel all $k||p_1$ singularities from ${\cal M}^{(2)}_{ss \to ss \to n}$ and the $n+2$ diagrams of ${\cal M}^{(2)}_{ss \to (sg)g \to n}$, for each permutation of the external momenta $\{q_1,\ldots q_n\}$. In fact, the same counterterm will also cancel all the collinear singularities of ${\cal M}^{(2)}_{ss \to ss \to n} + {\cal M}^{(2)}_{ss \to g(sg) \to n}$ as well as the soft singularities of all these diagrams.
 
In the other collinear limit $k \parallel p_2$ the factorization occurs analogously. We find
\begin{align}
    \lim_{k \parallel p_2} 
   {\cal M}_{n,{\rm IR}}^{(2)\text{,fact}} 
  =&-i\, g_s^2 \,\frac{C_A}{2}\, \frac{1}{k^2(k-p_2)^2}   \frac{(k-2p_2)\cdot p_1}{k\cdot p_1} \nonumber\\
 & \qquad \times
 \left( {\mathcal{M}}_n^{(1)}(l,p_1,p_2) + {\mathcal{M}}_n^{(1)}(l+k,p_1,p_2)\right),
 \label{eq:genampinkp2limit}
\end{align}
where we take $\chi=p_1$.
Finally, we note that the singular behavior in the soft limit, $k \sim 0$, comes entirely from the diagrams ${\cal M}^{(2)}_{ss \to ss \to n}$ and is equal to the soft limit of both Eq.~\eqref{eq:genampinkp1limit} and \eqref{eq:genampinkp2limit},
\begin{align}
\lim_{k \sim 0} \,  {\cal M}_{n,{\rm IR}}^{(2)\text{,fact}}
=-i\, g_s^2 \, C_A \, \frac{p_1 \cdot p_2 }{k^2 \, k\cdot p_1 \,k\cdot p_2} \times
  {\mathcal{M}}_n^{(1)}(l,p_1,p_2)\, .
 \label{eq:genampinksoftlimit}
\end{align}
We are now ready to identify the infrared counterterm that cancels these behaviors as in the first line of Eq.~\eqref{eq:GGsubtraction-termbyterm} for the hard-scattering function.

\subsection{The integrable hard function for \texorpdfstring{${\cal M}^{(2)\text{,fact}}_{n,\text{IR}}$}{Mfact}}

The factorized collinear and soft singularities of Eq.~\eqref{eq:genampinkp1limit}, \eqref{eq:genampinkp2limit} and Eq.~\eqref{eq:genampinksoftlimit}, which originate from the diagrams of Eq.~\eqref{eq:Mn2IRfact} are also present in a gauge theory of scalars. Motivated by this observation, we will use as an infrared counterterm an amplitude for a simple $2 \to 1$ scalar fusion process, which contains the same soft and collinear singularities as the diagrams of Eq.~\eqref{eq:Mn2IRfact}. This is a term of the form ${\cal F}^{(1)}_{\rm scalar}(k)\, {\cal M}^{(1)}_{gg \to {\rm colorless}}(l)$, as anticipated in Eq.~\eqref{eq:GGsubtraction} for the two-loop hard scattering function. 

In fact, the collinear and soft limits of ${\cal M}_{n,{\rm IR}}^{(2)\text{,fact}}$, exhibited in
Eqs.~\eqref{eq:genampinkp1limit} - \eqref{eq:genampinksoftlimit}
are matched by the product of a one-loop scalar form factor, which generates all infrared-singular behavior, multiplied
by the infrared finite Born amplitude, averaged over two values of loop momentum
\begin{eqnarray}
\label{eq:MirH-minus-FM}
{\cal H}_{n,{\rm IR}}^{(2)\text{,fact}}(k,l)
= {\cal M}_{n,{\rm IR}}^{(2)\text{,fact}}(k,l) -
{\cal F}_{\rm scalar}^{(1)}(k) \times  
\left< {\mathcal{M}}_n^{(1)}\right>_k\left( l,p_1,p_2\right) \, ,
\end{eqnarray}
where the short-distance function is given by
\begin{eqnarray}
    \left< {\mathcal{M}}_n^{(1)}\right>_k\left( l,p_1,p_2\right)
    \equiv \frac 1 2 \,  \left( {\mathcal{M}}_n^{(1)}(l,p_1,p_2) + {\mathcal{M}}_n^{(1)}(l+k,p_1,p_2)\right) \, .    
\end{eqnarray}
Note that this is not yet quite in the form for ${\cal H}_{n,{\rm IR}}^{(2)\text{,fact}}$ given in Eq.~\eqref{eq:GGsubtraction-termbyterm}, simply because $k$ and $l$ dependence is not fully separated.
The product $\times$ in Eq.~\eqref{eq:MirH-minus-FM} includes a sum over color indices, which is trivial because both ${\cal F}^{(1)}_{\rm scalar}$ and the one-loop function ${\cal M}_n^{1}$ are both diagonal in color.
In the following, we treat ${\cal F}_{\rm scalar}^{(1)}$ as a function, and associate the external color indices with those of the hard scattering. The scalar form factor in Eq.~\eqref{eq:MirH-minus-FM} is then defined by
\begin{eqnarray}
\label{eq:scalarformfactor}
    {\cal F}_{\rm scalar}^{(1)}(k) \, \delta_{ab} &=&  -i\, g_s^2\, C_A %\delta_{ab}\, 
    \frac{\left( k - 2 p_2\right) \cdot \left(k+ 2 p_1 \right) }{k^2 \, \left(k+ p_1\right)^2 \,\left(k - p_2\right)^2} 
    \, \delta_{ab}
    \nn\\ [2mm]
    &=& 
     \includegraphics[width=0.2\textwidth, page=2,valign=c]{figures_standalone/tikz_formfactor.pdf} \, ,
\end{eqnarray}
which can be given a graphical representation as shown. In the diagram, the external legs are truncated and carry no polarization vectors, which is indicated by the dots. 
The scalar-scalar vertex, depicted as a square, carries the color structure of the short-distance function, which is simply the Kronecker delta color tensor, $\delta_{cd}$ in the figure. 

We close this section by putting our integrable function ${\cal H}_{n,{\rm IR}}^{(2)\text{,fact}}$ into the form specified by Eq.~\eqref{eq:GGsubtraction-termbyterm}, as the sum of its diagrams minus a form factor subtraction times the one-loop amplitude and a shift subtraction. We do so by simply re-expressing our short distance function as
\begin{eqnarray}
    \left< {\mathcal{M}}_n^{(1)}\right>_k\left(l,p_1,p_2\right) 
    =  \mathcal{M}_n^{(1)} \left( l,p_1,p_2\right)  -
    \frac 1 2 \,  \left( {\mathcal{M}}_n^{(1)}(l,p_1,p_2) - {\mathcal{M}}_n^{(1)}(l+k,p_1,p_2)\right) \, ,   
    \nn\\
\end{eqnarray}
where clearly the integral over $l$ of the subtraction vanishes, as required by Eq.~\eqref{eq:bf-Delta-M-vanishes}.
This allows us to express Eq.~\eqref{eq:MirH-minus-FM} in the form %changed this to be correct
\begin{eqnarray}
\label{eq:MirH-minus-FM-Delta}
{\cal H}_{n,{\rm IR}}^{(2)\text{,fact}}(k,l)
&=& {\cal M}_{n,{\rm IR}}^{(2)\text{,fact}}(k,l) -
{\cal F}_{\rm scalar}^{(1)}(k) \times  
{\mathcal{M}}_n^{(1)}\left( l,p_1,p_2\right)
\nn\\[2mm]
&\ & + {\cal F}_{\rm scalar}^{(1)}(k) \times \frac{1}{2}
\left( {\mathcal{M}}_n^{(1)}(l,p_1,p_2) - {\mathcal{M}}_n^{(1)}(l+k,p_1,p_2)\right) \, ,
\end{eqnarray}
which matches Eq.~\eqref{eq:GGsubtraction-termbyterm} with 
\begin{align}
    \Delta {\cal M}_{n,{\rm IR}}^{(2),\text{fact}}(k,l) = {\cal F}_{\rm scalar}^{(1)}(k) \times \frac{1}{2}
\left(  {\mathcal{M}}_n^{(1)}(l+k,p_1,p_2)-{\mathcal{M}}_n^{(1)}(l,p_1,p_2)\right) \, .
\end{align}
Notice that now the form factor subtraction involves the standard one-loop amplitude, while the shift term is proportional to the difference of that amplitude at two values of its loop momentum. Although the short-distance function, the fermion loop here, is independent of the loop momentum $k$ of the form factor after integration over $l$, it retains all its non-local dependence on its external momenta and internal masses. The short-distance function thus contains in general the traces of arbitrary Dirac structures at fixed $l$. 
We again note that the subtractions necessary to provide ultraviolet convergence, which will be presented in Sec.~\ref{sec:UVCT}, do not change the pattern of these results.

This completes our construction of the integrable function, ${\cal H}_{n,{\rm IR}}^{(2)\text{,fact}}$, of the diagrams with factoring infrared and collinear singularities, ${\cal M}_{n,{\rm IR}}^{(2)\text{,fact}}$. We are now ready to turn to the diagrams of Eq.~\eqref{eq:ssfusdiag_generic_all_diags-0}, 
${\cal M}_{n,{\rm IR}}^{(2)\text{,shift}}$,
which are collinear-singular individually on the local level, but whose collinear singularities cancel after integration. As we shall see, these diagrams require only shift subtractions for local integrability.

\section{Shift-integrable collinear singularities}
\label{sec:NonLocalSings}

Continuing the discussion above we now consider the class of diagrams that are non-factorizable but shift-integrable, as discussed in Sec.~\ref{sec:shift-discussion},
\begin{align}
{\cal M}_{n,{\rm IR}}^{(2)\text{,shift}} 
    &\equiv 
\includegraphics[width=0.25\textwidth, page=2,valign=c]{figures_standalone/tikz_general_diagrams.pdf}
- \includegraphics[width=0.25\textwidth, page=9,valign=c]{figures_standalone/tikz_general_diagrams.pdf}
-\includegraphics[width=0.25\textwidth, page=18,valign=c]{figures_standalone/tikz_general_diagrams.pdf}
    \nonumber \\
   &  = \includegraphics[width=0.25\textwidth,
page=22,valign=c]{figures_standalone/tikz_general_diagrams.pdf} +   \includegraphics[width=0.25\textwidth, page=1,valign=c]{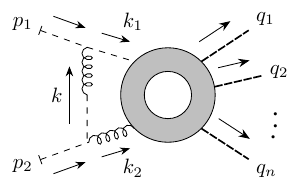} +  \includegraphics[width=0.25\textwidth, page=2,valign=c]{figures_standalone/tikz_diag_appendix.pdf}\nonumber \\
 & +\includegraphics[width=0.25\textwidth, page=3,valign=c]{figures_standalone/tikz_diag_appendix.pdf} +  \includegraphics[width=0.25\textwidth, page=4,valign=c]{figures_standalone/tikz_diag_appendix.pdf}  + \includegraphics[width=0.25\textwidth, page=5,valign=c]{figures_standalone/tikz_diag_appendix.pdf} \nonumber\\
&+\includegraphics[width=0.25\textwidth, page=6,valign=c]{figures_standalone/tikz_diag_appendix.pdf}
\label{eq:ssfusdiag_generic_all_diags}\, .
\end{align}
As above, we begin with the assignment of momentum flow for these diagrams, from which will follow the form of shift counterterms.
All diagrams in Eq.~\eqref{eq:ssfusdiag_generic_all_diags} include a propagator which directly connects the incoming states $p_1$ and $p_2$. We assign to this propagator line a momentum label $k$, which we direct from $p_2$ to $p_1$. The loop momenta which are external and incoming to the fermion-loop subgraphs are then 
\begin{equation}
k_1 = k+p_1, \quad k_2 = p_2 -k.
\end{equation}
This choice of momentum flow results in a manifestly symmetric expression for the integrand under the exchange $(\pol_1, p_1) \leftrightarrow (\pol_2, - p_2)$ and will lead conveniently to symmetric shift counterterms for collinear singularities. 
We will also need to assign a label $l$ for the internal loop momentum of the fermion loop. We need not specify the momentum $l$ yet, but, later, we will make a specific choice for $l$ that will lead to an appealing graphical interpretation of the counterterms. This choice will be different than the assignment that we made for ${\cal M}_{n,{\rm IR}}^{(2)\text{,fact}}$ in Sec.~\ref{sec:momflowsM2fact}.
In what follows, index $\alpha$ ($\beta$) will label the vertex at which the line $k_1$ ($k_2$) attaches to the fermion loop, as in the QED Ward identities of Eq.~\eqref{eq:WIloopAsec3} to which we will appeal. 

The sum of diagrams in Eq.~\eqref{eq:ssfusdiag_generic_all_diags} can be written in a convenient form by eliminating terms with $p_1^\alpha$ in favor of $k^\alpha$ and $k_1^\alpha$, and similarly for $p_2^\beta$. 
After straightforward algebra, we can write the result of this procedure as
\begin{align}
{\cal M}_{n,{\rm IR}}^{(2)\text{,shift}} = &i\, g_s^2 \, \delta_{ab}  C_A
\frac{ \widetilde{\mathcal{M}}^{(1)}_{n\,\alpha \beta}\left(l,k_1, k_2 \right)}{ k^2 \, k_1^2 \, k_2^2} \,\nonumber \\
&
\times
\Big( U^{\alpha\beta} + 4 \left[
-k^\alpha\, k^\beta \, \epsilon_1 \cdot \epsilon_2
+k^\alpha\,  \epsilon_1^\beta \, \epsilon_2 \cdot k
+k^\beta \, \epsilon_2^\alpha \, \epsilon_1 \cdot k
\right.
\nonumber \\
&\left.
\hspace{1.1cm} + \epsilon_1^\alpha \Big(p_1^\beta \, \epsilon_2\cdot k
- k^\beta \, \epsilon_2\cdot p_1 \Big)
- \epsilon_2^\beta \Big(p_2^\alpha\,  \epsilon_1\cdot k
- k^\alpha\,  \epsilon_1\cdot p_2 \Big)
\right]\Big).
\label{eq:ssfusdiag_generic_all}
\end{align}
The fermion-loop subgraphs $\widetilde{\mathcal{M}}^{(1)}_{n\,\alpha \beta}$ include all diagrams of the Born amplitude and are defined in Eq.~\eqref{eq:Mtilde}, with polarization vectors and a color tensor removed. As above, $l$ is the internal loop momentum of the $\widetilde{\mathcal{M}}^{(1)}_{n\,\alpha \beta}$ subgraphs. 
All terms with $k_1^\alpha$ and/or $k_2^\beta$, that is, with scalar-polarized lines attached to the fermion loop, are collected in the tensor $U^{\alpha\beta}$, which is given by
\begin{align}
    U^{\alpha\beta} & = \epsilon_1 \cdot \epsilon_2 \,
    k_1^\alpha k_2^\beta  
+k_1^\alpha \Big( 
2 k^\beta \, \epsilon_1 \cdot \epsilon_2 
- 2 \epsilon_1^\beta \, k \cdot \epsilon_2 
+\epsilon_2^\beta \, (k-2 p_2) \cdot \epsilon_1
\Big)
\nonumber\\ 
& \hspace{1cm}
+ k_2^\beta \Big(-2 k^\alpha \, \epsilon_1 \cdot \epsilon_2 
+ 2 \epsilon_2^\alpha \, k \cdot \epsilon_1 
-\epsilon_1^\alpha \, (k+2 p_1) \cdot \epsilon_2
\Big).
\label{eq:ssfus_U}
\end{align}
Regarding the remaining terms of Eq.~\eqref{eq:ssfusdiag_generic_all}, the first term in the squared bracket results from the first diagram in Eq.~\eqref{eq:ssfusdiag_generic_all_diags-0}, $\mathcal{M}^{(2)}_{ss\to gg\to n}$ (equivalently, the first diagram of the second equality in \eqref{eq:ssfusdiag_generic_all_diags}). 
The second and third terms in Eq.~\eqref{eq:ssfusdiag_generic_all} correspond to the fifth and sixth diagrams of \eqref{eq:ssfusdiag_generic_all_diags-0}, $\mathcal{M}^{(2)}_{sg\to gs\to n}$ and $\mathcal{M}^{(2)}_{gs\to sg\to n}$. On the last line, the first combination of terms belongs to the second and fourth diagrams of \eqref{eq:ssfusdiag_generic_all_diags-0}, $\mathcal{M}^{(2)}_{ss\to sg\to n}$ and $\mathcal{M}^{(2)}_{sg\to ss\to n}$, which vanish in the collinear limit $k \parallel p_1$. The two terms in the last bracket of Eq.~\eqref{eq:ssfusdiag_generic_all} originate from the third and seventh diagram in Eq.~\eqref{eq:ssfusdiag_generic_all_diags-0}, $\mathcal{M}^{(2)}_{ss\to gs\to n}$ and $\mathcal{M}^{(2)}_{gs\to ss\to n}$. This combination analogously vanishes in the collinear limit $k \parallel p_2$.

As discussed in Section~\ref{sec:IRsing_channels}, there are only two singular regions that produce infrared singularities: the region where $k_1$ (or $k$) is collinear to $p_1$ and where $k_2$ (or $k$) is collinear to $p_2$. The two regions intersect at a single point, at $k=k_1-p_1=p_2-k_2=0$, at which the exchanged vertical line between the scalar and gluons seen in Eq.~\eqref{eq:ssfusdiag_generic_all} carries vanishing momentum, i.e., $k\to 0$. The other soft limits, where $k_1$ or $k_2$ carry zero momentum, are fully disconnected, since $k_1=0$ implies $k_2=\sum_{i=1}^n q_i$, and vice-versa. At these endpoints, the integrals are power-counting finite. Examining the explicit expressions for the integrand of ${\cal M}_{n,{\rm IR}}^{(2)\text{,shift}}$ above, we see that all soft singularities are in the term with $U^{\alpha\beta}$, since every other term has at least one factor of the vector $k^\mu$ in the numerator.

We note that with a special choice of polarization vectors satisfying
\begin{align}
\label{eq:polarisations_ss}
\pol_1 \cdot p_2 = \pol_2 \cdot p_1 =0 \, ,
\end{align}
only the first diagram in Eq.~\eqref{eq:ssfusdiag_generic_all_diags} and therefore only terms proportional to $\epsilon_1\cdot \epsilon_2$ in Eqs.~\eqref{eq:ssfusdiag_generic_all} and \eqref{eq:ssfus_U} remain infrared singular. This choice would simplify the construction of counterterms. In view of a future generalization of our method to processes with more than two external gluons, however, it may prove useful to make choices other than the one in Eq.~\eqref{eq:polarisations_ss}, which is why we carry on here in a fully general manner.

\subsection{Subtractions for \texorpdfstring{${\cal M}_{n,{\rm IR}}^{(2)\text{,shift}}$}{Mshift}: eliminating soft divergences}

As we will review below, all terms in Eq.~\eqref{eq:ssfusdiag_generic_all} that have a factor $k_1^\alpha$ or $k_2^\beta$, that is, all terms in $U^{\alpha\beta}$, vanish by the QED Ward identity, 
Eq.~\eqref{eq:WIloopAsec3},
\begin{align}
\label{eq:WIloopA}
    \int d^D l\; k_1^{\alpha} \, {\widetilde {\cal M}}^{(1)}_{n\,\alpha\beta}\left(l,k_1, k_2\right) = \int \mathrm{d}^D l\; k_2^{\beta} \,  {\widetilde {\cal M}}^{(1)}_{n\,\alpha\beta}\left( l,k_1, k_2\right) = 0 \, ,
\end{align}
which states that in QED fermion loops are gauge invariant after integration.
Thus, a scalar-polarized vector decouples from a fermion loop in QED, for arbitrary polarizations of additional vector lines. This is not generally the case for a nonabelian fermion loop.

We can use the QED Ward identity on gluons in the shift-integrable amplitude, Eq.~\eqref{eq:ssfusdiag_generic_all}, because the color factors of all diagrams in $\widetilde{\cal M}^{(1)}_{n,\alpha\beta}$ are equal.
Terms that represent the sum over diagrams with scalar polarizations thus satisfy the condition for shift subtractions in Eq.~\eqref{eq:bf-Delta-M-vanishes}. We can therefore subtract them from the amplitude without changing its value after integration.
In fact, in the following we will exploit the Ward identities of Eq.~\eqref{eq:WIloopA} in order to eliminate locally all the remaining singularities of the individual diagrams in Eq.~\eqref{eq:ssfusdiag_generic_all}. 

To begin this process, we subtract the terms manifestly proportional to $k_1^\alpha$ or $k_2^\beta$, hence the entire term $ U^{\alpha\beta}$. In the notation of Eq.~\eqref{eq:bf-M-def}, we find an equivalent integrand without soft singularities, although with collinear singularities remaining. We denote this integrand as
\begin{eqnarray}
\label{eq:ssfusdiag_nosoft}
{\widehat {\cal M}}^{(2)\text{,shift}}_{n\text{,IR,}\bar{U}}  &=&
{\cal M}^{(2)\text{,shift}}_{n\text{,IR}}
- i\, g_s^2 \, \delta_{ab}  C_A
\frac{ \widetilde{\mathcal{M}}^{(1)}_{n\,\alpha \beta}\left(l,k_1, k_2 \right)}{ k^2 \, k_1^2 \, k_2^2} U^{\alpha\beta}
\nn\\[2mm]
&=&
{\cal M}^{(2)\text{,shift}}_{n\text{,IR}} - \Delta_U {\cal M}^{(2)\text{,shift}}_{n,\text{IR,}U}
\nn\\[2mm]
&=&  4\,i\, g_s^2 \, \delta_{ab}  C_A
\frac{ \widetilde{\mathcal{M}}^{(1)}_{n\,\alpha \beta}\left(l,k_1, k_2 \right)}{ k^2 \, k_1^2 \, k_2^2} 
 \left[
-k^\alpha\, k^\beta \, \epsilon_1 \cdot \epsilon_2
+k^\alpha\,  \epsilon_1^\beta \, \epsilon_2 \cdot k
+k^\beta \, \epsilon_2^\alpha \, \epsilon_1 \cdot k
\right.
\nonumber \\
&\ &\left.
\hspace{1.1cm} + \epsilon_1^\alpha \Big(p_1^\beta \, \epsilon_2\cdot k
- k^\beta\,  \epsilon_2\cdot p_1 \Big)
- \epsilon_2^\beta \Big(p_2^\alpha\,  \epsilon_1\cdot k
- k^\alpha \, \epsilon_1\cdot p_2 \Big)
\right]\, ,
\end{eqnarray}
where the second equality defines our first shift subtraction, as simply all the terms of $U^{\alpha\beta}$.
As is seen from the third equality, every term remaining in the numerator after the subtraction of $\Delta_U {\cal M}$ vanishes in the soft limit, which renders the soft region integrable. The integrand of ${\widehat {\cal M}}^{(2)\text{,shift}}_{n\text{,IR,}\bar{U}}$, however, is still singular in collinear limits. 

To treat the collinear singularities that remain in ${\widehat {\cal M}}^{(2)\text{,shift}}_{n\text{,IR,}\bar{U}}$, we will use the same Ward identities to show that the collinear singularities of the integrand in Eq.~\eqref{eq:ssfusdiag_nosoft} entirely cancel after the $l$ integration is carried out. To remove these remaining singularities, we need only construct additional shift subtractions, which, as usual, leave the integrated amplitude unchanged. We will see that these shift subtractions can be given a useful graphical interpretation, which we believe will facilitate their implementation in numerical evaluations.

\subsection{Subtractions for \texorpdfstring{${\cal M}_{n,{\rm IR}}^{(2)\text{,shift}}$}{Mshiftsub}: eliminating collinear divergences}

Again recalling the notation of Eq.~\eqref{eq:bf-Delta-M-vanishes}, we denote shift subtractions for collinear singularities of ${\widehat {\cal M}}^{(2)\text{,shift}}_{n\text{,IR,}\bar{U}}$ by $\Delta_1 \,  {\cal M}^{(2)\text{,shift}}_{n\text{,IR,}\bar{U}}$ and $ \Delta_2 \,  {\cal M}^{(2)\text{,shift}}_{n\text{,IR,}\bar{U}} $, 
which will be constructed to remove collinear singularities 
in the $k \parallel p_1$ and $k \parallel p_2$ limits, respectively. This will bring us to the locally-integrable hard function
${\cal H}^{(2)\text{,shift}}_{n\text{,IR}}$ of Eqs.~\eqref{eq:calH-threeterms} and \eqref{eq:GGsubtraction-termbyterm}, in the form
\begin{eqnarray}
\label{eq:sstoggton_bold}
{\cal H}^{(2)\text{,shift}}_{n\text{,IR}}    
&\equiv& 
{\widehat {\cal M}}^{(2)\text{,shift}}_{n\text{,IR,}\bar{U}}
-  \Delta_1 \, {\cal M}^{(2)\text{,shift}}_{n\text{,IR,}\bar{U}} 
-\Delta_2 \, {\cal M}^{(2)\text{,shift}}_{n\text{,IR,}\bar{U}}    \nn\\[2mm]
&=&  {\cal M}^{(2)\text{,shift}}_{n\text{,IR}} - \Delta_U {\cal M}^{(2)\text{,shift}}_{n,\text{IR,}U}
-  \Delta_1 \, {\cal M}^{(2)\text{,shift}}_{n\text{,IR,}\bar{U}} 
-\Delta_2 \, {\cal M}^{(2)\text{,shift}}_{n\text{,IR,}\bar{U}}    \, ,
\nn\\[2mm]
&=&  {\cal M}^{(2)\text{,shift}}_{n\text{,IR}} - \Delta {\cal M}^{(2)\text{,shift}}_{n,\text{IR,}}
\end{eqnarray}
where in the second equality, ${\widehat {\cal M}}^{(2)\text{,shift}}_{n\text{,IR,}\bar{U}}$ has been re-expressed using Eq.~\eqref{eq:ssfusdiag_nosoft}, to put the integrable function into the notation of Eq.~\eqref{eq:GGsubtraction-allterms}, and in the third equality introduced the notation of Eq.~\eqref{eq:GGsubtraction-termbyterm} for the complete shift.

We now show that we can construct $\Delta_1 \cal M$ and $\Delta_2 \cal M$ counterterms so that they are proportional to either $k_1^\alpha$ or $k_2^\beta$, respectively, allowing the implementation of the 
QED Ward identities, Eq.~\eqref{eq:WIloopA}. 
To motivate their construction below, we examine the collinear limit $k \parallel p_1$ of Eq.~\eqref{eq:ssfusdiag_nosoft}, finding that only a few terms survive unsuppressed by the numerator,
\begin{align}
   \lim_{k = -x p_1}  {\widehat {\cal M}}^{(2)\text{,shift}}_{n \text{,IR,}\bar{U}}
&= 4\, i\, g_s^2 \, \delta_{ab}  C_A
\frac{ k^\alpha\, \widetilde{\mathcal{M}}^{(1)}_{n\,\alpha \beta}\left(l,k_1, k_2 \right)}{ k^2 \, k_1^2 \, k_2^2} \,\nonumber \\
&\hspace{1cm}
\times
\,\left(
- k^\beta \, \epsilon_1 \cdot \epsilon_2
+ \epsilon_1^\beta \, \epsilon_2 \cdot k + \epsilon_2^\beta \,\epsilon_1\cdot p_2
\right).
\end{align}
Of the seven terms in Eq.~\eqref{eq:ssfusdiag_nosoft} the two numerator terms proportional to $\epsilon_1^\alpha$ cancel in the $k \parallel p_1$ limit, while two are proportional to $\epsilon_1\cdot k$, which vanishes in this limit. This leaves the three terms shown.

For the corresponding subtraction, we recast this collinear approximation into a form in which the gluon $k_1$ attached to the fermion loop is exactly scalar-polarized, introducing into
the expression an auxiliary vector $\xi_1^\mu$, 
\begin{align}
\label{eq:T1coll}
\Delta_1 \, {\cal M}^{(2)\text{,shift}}_{n\text{,IR,}\bar{U}}
= &\, i\, g_s^2 \, \delta_{ab} \,  C_A 
\dfrac{k_1^\alpha\, \widetilde{\mathcal{M}}^{(1)}_{n\,\alpha \beta}\left(l,k_1, k_2 \right) }{ k^2 \, k_1^2 }\nonumber\\
&\times 
\dfrac{2\,  p_1 \cdot (2 \xi_1 -k_1) \left(
- k^\beta \, \epsilon_1 \cdot \epsilon_2
+ \epsilon_1^\beta \, \epsilon_2 \cdot k + \epsilon_2^\beta \,\epsilon_1\cdot p_2
\right) 
}{p_1 \cdot p_2 \, \left( k_1^2 - 2 k_1 \cdot \xi_1 \right)} \, .
\end{align}
At this stage, we only require that $p_1\cdot \xi_1,\xi_1^2\ne 0$, but we will see that a specific choice for $\xi_1$ will give the subtraction an appealing diagrammatic interpretation. For any $\xi_1$, this expression provides a singular integral only when $k$ is collinear to $p_1$; it is finite in both the soft and $p_2$-collinear regions.

For the other collinear limit, we take, correspondingly, 
\begin{align}
\label{eq:T2coll}
    \Delta_2 \, {\cal M}^{(2)\text{,shift}}_{n\text{,IR,} \bar{U}} 
=& \, i\, g_s^2 \,\delta_{ab} \,  C_A 
\frac{k_2^\beta\, \widetilde{\mathcal{M}}^{(1)}_{n\,\alpha \beta}\left(l,k_1, k_2 \right)}{k^2 \, k_2^2 } \nonumber\\
&\times \frac{-2 \, p_2 \cdot (2 \xi_2 -k_2)\left(
- k^\alpha \, \epsilon_1 \cdot \epsilon_2
+ \epsilon_2^\alpha \, \epsilon_1 \cdot k -\epsilon_1^\alpha \,\epsilon_2\cdot p_1
\right)  
}{p_1 \cdot p_2 \,\left(  k_2^2 - 2 k_2 \cdot \xi_2 \right)} ,
\end{align}
in terms of an auxiliary vector $\xi_2^\mu$, satisfying $p_2 \cdot \xi_2,\xi_2 ^2 \neq 0$. 

It is easy to verify with an explicit computation that in the collinear limits, for $k= -x p_1$ or $k = x p_2$, the combination on the right-hand side of Eq.~\eqref{eq:sstoggton_bold} is integrable. In the next section we will construct a new version of these counterterms, which both makes their vanishing after the $l$ loop momentum integration manifest by casting them as shift counterterms, and which, with an appropriate choice of the vectors $\xi_i$, gives them graphical interpretations.

\subsection{The one-loop Ward identity and shift counterterms}

\label{sec:one-loop_WI}
We will now recall the standard diagrammatic demonstration of the one-loop Ward identity of Eq.~\eqref{eq:WIloopA}. Although the result is a standard one, the demonstration provides in parallel the necessary ingredients for the construction of the collinear counterterms of Eqs.~\eqref{eq:T1coll} - \eqref{eq:T2coll} on the right-hand side of Eq.~\eqref{eq:sstoggton_bold}. It will also enable us to introduce a convenient diagrammatic construction of the collinear shift counterterms, utilizing the freedom to choose the auxiliary vectors $\xi_i$. 

The one-loop tensor ${\widetilde {\cal M}}^{(1)}_{n\,\alpha\beta}\left( l,k_1, k_2\right)$ can be written as
\begin{align}
\label{eq:Mtilde2Lcal}
{\widetilde {\cal M}}^{(1)}_{n\,\alpha\beta}\left( l,k_1, k_2\right)
 &= %\delta_{ab}\, 
 \sum_{{\rm perms}\, \{q_{n-1},\dots , q_1\}} {\cal L}_{\alpha\beta}^{(n+2)}  \left (l,k_1,k_2,q_n\dots q_1\right)\, ,
\end{align}
where 
we recall from Eq.~\eqref{eq:ssfusdiag_generic_all} that its color tensor, $\delta_{ab}$, is suppressed in the definition of $\widetilde{\cal M}^{(1)}_{n,\alpha\beta}$.
The full set of diagrams that contribute to the loop function ${\cal L}_{\alpha\beta}$ is given by  
\begin{align}
\label{eq:LabexplicitA}
    & \, {\cal L}_{\alpha\beta}^{(n+2)} \left (l,k_1,k_2,q_n\dots q_1\right)\delta_{ab} \nn\\
    &= \sum_{m=0}^{n-1} \sum_{r=0}^{m} \left( \includegraphics[width=0.4\textwidth, page=1,valign=c]{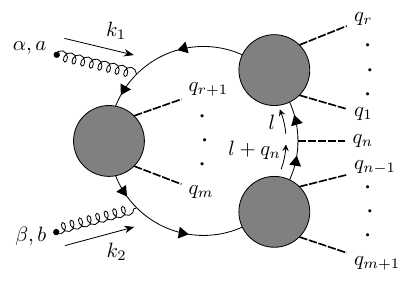}  + \includegraphics[width=0.4\textwidth, page=2,valign=c]{figures_standalone/tikz_ssfusion.pdf}\right). \nn\\
\end{align}
In the above graphs, the incoming gluons are truncated, indicated by the dots, and contribute the correct color and Lorentz indices to ${\widetilde {\cal M}}^{(1)}_{n\,\alpha\beta}\left( l,k_1, k_2\right)$.
To specify the loop momentum $l$ for this set of diagrams, we choose, arbitrarily, one of the Higgs boson momenta, labeled $q_n$, to play a special role in our momentum assignments. Other choices are possible, and can lead to alternative shift counterterms.

In each diagram the fermion line whose arrow flows into the vertex from which momentum $q_n$ emerges is defined to carry momentum $l + q_n$, so that the line flowing out carries momentum $l$. Note that the assignment of the momentum label $l$ here is different than the choice that we made for ${\cal M}_{n,{\rm IR}}^{(2)\text{,fact}}$ in Sec.~\ref{sec:momflowsM2fact}. To specify the momenta of other lines explicitly, we introduce the notation
\begin{eqnarray}
    Q_a^b = \sum_{i=a}^b q_i\,, \quad \text{with }\, b>a,
    \label{eq:Qab-def}
\end{eqnarray}
in terms of which, from momentum conservation,
\begin{eqnarray}
    Q_1^n = k_1 + k_2\, .
    \label{eq:modmentum-conservation}
\end{eqnarray}
In this notation, we have, for example, 
\begin{eqnarray}
      k_1+k_2 - Q_1^m %Q_m^n 
      = Q_{m+1}^n\, , 
    \label{eq:Q-relation}
\end{eqnarray}
which we will use shortly in the diagrams.

The full set of diagrams naturally separates into two subsets, those in which $k_1$ appears before $k_2$ as we follow the arrow from the $q_n$ vertex, and those in which $k_2$ proceeds $k_1$. The sum over permutations in Eq.~\eqref{eq:Mtilde2Lcal} refers to the ordering of the remaining Higgs boson vertices (after we single out the special reference vertex with momentum $q_n$) along the direction of charge flow in the fermion loop, up to cyclic permutations.

Let us now consider the contraction of the fermion loop diagrams in ${\cal L}_{\alpha\beta}^{(n+2)}$ with a longitudinal polarization. Applying the Ward identity of Eq.~\eqref{eq:WI-blob} to the diagrams in Eq.~\eqref{eq:LabexplicitA}, we find four terms
\begin{align}
&k_1^{\alpha}\,{\cal L}_{\alpha\beta}^{(n+2)}  \left (l,k_1,k_2,q_n\dots q_1\right) \delta_{ab} \nn\\
     &= \sum_{m=0}^{n-1} 
     \includegraphics[width=0.44\textwidth, page=4,valign=c]{figures_standalone/tikz_ssfusion.pdf} +  \includegraphics[width=0.46\textwidth, page=3,valign=c]{figures_standalone/tikz_ssfusion.pdf}
     \nn\\
     &= \sum_{m=0}^{n-1} 
    \raisebox{2.4mm}{\includegraphics[width=0.42\textwidth, page=5,valign=c]{figures_standalone/tikz_ssfusion.pdf}} +  \includegraphics[width=0.42\textwidth, page=6,valign=c]{figures_standalone/tikz_ssfusion.pdf} \nn\\
     & \qquad 
     + \includegraphics[width=0.44\textwidth, page=7,valign=c]{figures_standalone/tikz_ssfusion.pdf} +  \raisebox{3mm}{\includegraphics[width=0.48\textwidth, page=8,valign=c]{figures_standalone/tikz_ssfusion.pdf}}
\label{eq:k1-wi-full-loop-A}
\end{align}
We indicate the truncated line still with the dot and the additional $k_1^\alpha$ term with the dot-dashed arrow. The second and the third diagrams on the right-hand side of Eq.~\eqref{eq:k1-wi-full-loop-A} cancel each other, since, as we can see using Eq.~\eqref{eq:Q-relation}, these are exactly the same diagrams with a different overall sign (due to the rules of Eq.~\eqref{eq:gqqvertexWard}). We are therefore left with 
\begin{align}
\label{eq:WIloopResult1}
    &k_1^{\alpha}\,{\cal L}_{\alpha\beta}^{(n+2)}  \left (l,k_1,k_2,q_n\dots q_1\right) \delta_{ab}\nn\\
    &= \sum_{m=0}^{n-1} 
    \includegraphics[width=0.42\textwidth, page=5,valign=c]{figures_standalone/tikz_ssfusion.pdf} +  \includegraphics[width=0.475\textwidth, page=8,valign=c]{figures_standalone/tikz_ssfusion.pdf}.\nn\\
\end{align}
Let us now define  
\begin{eqnarray}
\delta_{ab} \, {\cal L}^{(n+1)}_\beta\left( l,k_2,q_n\dots q_1 \right) \equiv 
\sum_{m=0}^{n-1}   \includegraphics[width=0.42\textwidth, page=5,valign=c]{figures_standalone/tikz_ssfusion.pdf} 
\end{eqnarray}
to be the first sum on the right-hand side of Eq.~\eqref{eq:WIloopResult1}, consisting of fermion loops with $n+1$ lines and vertices, with a single external gluon, of momentum $k_2$. 
With a reflection of the diagrams to point the fermion arrow to a clockwise direction and using the rules of Eq.~\eqref{eq:gqqvertexWard} and the identity \eqref{eq:Q-relation}, we can see that the second sum on the right-hand side of Eq.~\eqref{eq:WIloopResult1} is the same function with an opposite sign and the $l$ loop momentum shifted to $l+k_1$. Therefore Eq.~\eqref{eq:WIloopResult1} becomes
\begin{align}
\label{eq:WIloopResult2}
     k_1^{\alpha}\,{\cal L}_{\alpha\beta}^{(n+2)}  \left (l,k_1,k_2,q_n\dots q_1\right)
= &{\cal L}^{(n+1)}_\beta\left( l,k_2,q_n\dots q_1 \right)\nn\\
& \hspace{1cm}-{\cal L}^{(n+1)}_\beta\left( l+k_1,k_2,q_n\dots q_1 \right)\, .
\end{align}
This makes the vanishing of the integrals in Eq.~\eqref{eq:WIloopA} manifest. It also shows that the subtractions of terms like $\Delta_i \, {\cal M}^{(2)\text{,shift}}_{n\text{,IR,}\bar{U}}$, Eqs.~\eqref{eq:T1coll} and \eqref{eq:T2coll}, 
which vanish by the QED Ward identity, are rightly classed as shift subtractions. 

We can now apply Eq.~\eqref{eq:WIloopResult2} to our proposed shift subtractions (counterterms), Eqs.~\eqref{eq:T1coll} and \eqref{eq:T2coll}, to derive shift subtractions with a natural graphical representation.
Choosing the auxiliary vector $\xi^\mu_1$ to equal $q_n^\mu$ in \eqref{eq:T1coll}, for example, we find a subtraction that has only quadratic denominators and the same collinear limit as in Eq.~\eqref{eq:ssfusdiag_nosoft}. 
 
Applying the Ward identity of Eq.~\eqref{eq:WIloopResult2} to ${\widetilde {\cal M}}^{(1)}_{n\,\alpha\beta}\left( l,k_1, k_2\right)$ in Eq.~\eqref{eq:T1coll} with $\xi^\mu_1=q_n^\mu$, for each of the terms that make up the one-loop amplitude in Eq.~\eqref{eq:Mtilde2Lcal}, the shift subtraction for $ {\cal M}^{(2)\text{,shift}}_{n\text{,IR,}\bar{U}}$ becomes 
\begin{align}
\label{eq:T1collphys_diag1}
    \Delta_1 \, {\cal M}^{(2)\text{,shift}}_{n\text{,IR,}\bar{U}}
&= i\, g_s^2 \,  C_A
\frac{2 p_1^\alpha \, \left(
 -k^\beta \, \epsilon_1 \cdot \epsilon_2
+ \epsilon_1^\beta \, \epsilon_2 \cdot k + \epsilon_2^\beta \,\epsilon_1\cdot p_2
\right)}{ p_1 \cdot p_2\, k^2 \, k_1^2 }  
\nonumber \\ 
&
\hspace{-3.5cm}
 \times \sum_{{\rm perms}\, \{q_{n-1}\dots q_1\} } 
 \left[
  \delta_{ab} \frac{\left(2 q_n - k_1 \right)_\alpha }{  k_1^2 - 2 k_1 \cdot q_n }
 {\cal L}^{(n+1)}_\beta\left( l,k_2,q_n\dots q_1 \right) 
- \left( l \to l+k_1 \right)
\right].
\end{align}
Again, the use of the Ward identity results in a shift counterterm. We can represent the terms in the bracket diagrammatically as 
\begin{align}
&\delta_{ab}\frac{\left(2 q_n - k_1 \right)_\alpha }{  k_1^2 - 2 k_1 \cdot q_n }
 {\cal L}^{(n+1)}_\beta\left( l,k_2,q_n\dots q_1 \right)
=\sum_{m=0}^{n-1} 
\includegraphics[width=0.45\textwidth, page=9,valign=c]{figures_standalone/tikz_ssfusion.pdf},
\label{eq:k1-subtraction-A}
\end{align}
where we have introduced a fictitious colored scalar-gluon-Higgs vertex with Feynman rule
\begin{align}
\includegraphics[width=0.2\textwidth, page=10,valign=c]{figures_standalone/tikz_ssfusion.pdf}
\equiv  \delta_{ab} (2 q_n - k_1)^\alpha  .
\end{align}
The dot again denotes that the gluon is truncated, that is, we have no propagator and no polarization vector. It only carries color and the Lorentz index and inserts the momentum $k_1$. In doing so, we have traded one of the fermion lines for an external final-state scalar line. The subtraction now takes the graphical form
\begin{align}
\label{eq:T1collphys_diag2}
    \Delta_1 \, {\cal M}^{(2)\text{,shift}}_{n\text{,IR,}\bar{U}}
=& i\, g_s^2 \,  C_A
\frac{2 p_1^\alpha \, \left(-
 k^\beta \, \epsilon_1 \cdot \epsilon_2
+ \epsilon_1^\beta \, \epsilon_2 \cdot k +\epsilon_2^\beta \,\epsilon_1\cdot p_2
\right)}{ p_1 \cdot p_2\, k^2 \, k_1^2 }  
\nonumber \\ 
& 
\hspace{-1cm}
 \times \sum_{{\rm perms}\, \{q_{n-1}\dots q_1\} } \left[
\includegraphics[width=0.43\textwidth, page=9,valign=c]{figures_standalone/tikz_ssfusion.pdf}
- \left( l \to l+k_1 \right)
\right].\nn\\
\end{align}
Comparing the collinear approximation of Eq.~\eqref{eq:T1collphys_diag2} and the amplitude in Eq.~\eqref{eq:ssfusdiag_generic_all} we notice that in the collinear counterterm the $k_2^2$ denominator has been eliminated in favor of a fictitious vertex on an outgoing colorless particle. To generate the proper color factor, we associate a generator $T_a$ in fundamental representation with the vertex at which the fictitious colored scalar with momentum $q_n-k_1$ attaches to the fermion loop, as in Eq.\ \eqref{eq:scalar-to-fermion}.

Given the symmetric assignments of momenta we are using, it is clear that an analogous subtraction for the single-collinear region, $k_2||p_2$, is given by 
\begin{eqnarray}
\label{eq:T2collphys}
    \Delta_2 \, {\cal M}^{(2)\text{,shift}}_{n\text{,IR,}\bar{U}}
&=& i\, g_s^2 \, C_A \frac{-2\, p_2^\beta\left(-\epsilon_1\cdot\epsilon_2 \,k^\alpha + \epsilon_1\cdot k\,\epsilon_2^\alpha -\epsilon_1^\alpha \,\epsilon_2\cdot p_1 \right)
}{p_1 \cdot p_2 \, k^2 \, k_2^2 }  
\nonumber \\ 
&& 
\hspace{-3.5cm}
 \times \sum_{{\rm perms}\, \{q_{n-1}\dots q_1\} } 
 \left[
  \delta_{ab} \frac{\left(2 q_n - k_2 \right)_\beta }{  k_2^2 - 2 k_2 \cdot q_n }
 {\cal L}^{(n+1)}_\alpha\left( l,k_1,q_n\dots q_1 \right) 
- \left( l \to l+k_2 \right)
\right],
\end{eqnarray}
where the propagator $k_1^2$ has been eliminated. The subtraction takes the graphical form
\begin{align}
\label{eq:T2collphys_diag2}
    \Delta_2 \, {\cal M}^{(2)\text{,shift}}_{n\text{,IR,}\bar{U}}
=&  i\, g_s^2 \, C_A \frac{-2\, p_2^\beta\left(-\epsilon_1\cdot\epsilon_2 \,k^\alpha + \epsilon_1\cdot k\,\epsilon_2^\alpha -\epsilon_1^\alpha \,\epsilon_2\cdot p_1 \right)
}{p_1 \cdot p_2 \, k^2 \, k_2^2 }  
\nonumber \\ 
&
\hspace{-1cm}
 \times \sum_{{\rm perms}\, \{q_{n-1}\dots q_1\} } 
 \left[
\includegraphics[width=0.43\textwidth, page=11,valign=c]{figures_standalone/tikz_ssfusion.pdf}
- \left( l \to l+k_2 \right)
\right].\nn\\
\end{align}
We close this discussion by noting again that our shift counterterm construction is by no means unique, and that it depends in particular on our choices for the flow of loop momenta.

\subsection{A class of shift-integrable singularities in quark-antiquark annihilation}

The method that we developed for the shift-integrable diagrams of Eq.~\eqref{eq:ssfusdiag_generic_all} can also be applied to an analogous class of Feynman diagrams that appear in electroweak production processes from quark annihilation. A local subtraction method for the latter was proposed earlier in Ref.~\cite{Anastasiou:2020sdt}. The alternative method that we propose here uses simpler counterterms, with fewer factors in their denominators.

The quark-fusion diagrams are found from the first diagram on the right-hand side of the second equality in Eq.~\eqref{eq:ssfusdiag_generic_all_diags}, simply by replacing the scalar lines by fermion lines. The integrand has the form 
\begin{align}
\label{eq:qqbarfusdiag}
    \mathcal{M}^{(2)}_{q\bar q \to gg \to  n} &= \includegraphics[width=0.35\textwidth, page=12,valign=c]{figures_standalone/tikz_ssfusion.pdf} 
    \nonumber \\ 
    &
    = -i \, g_s^2 \,\delta_{ij}\, C_F \,  \frac{{\bar v}(p_2)\gamma^\beta  \slashed{k} \gamma^\alpha u(p_1) }{k^2 k_1^2 k_2^2}  \, \widetilde{\mathcal{M}}^{(1)}_{n\,\alpha \beta}\left(l,k_1, k_2 \right),
\end{align}
where $i$ and $j$ are the quark color indices. The integrands represented by this expression are already finite in the soft limit $k \to 0$ due to a suppression from the $\slashed k$ factor in the numerator. Once again, however, the integrands of this type are singular in the collinear limits $k \parallel p_1$ and $k \parallel p_2$. As discussed above, we choose $q_n$ to play a special role in the momentum assignment, and we can write the colorless one-loop amplitude without polarization vectors as in Eqs.~\eqref{eq:Mtilde2Lcal} and \eqref{eq:LabexplicitA}. Given the Ward identities of Eq.~\eqref{eq:WIloopA}, we can remove these collinear singularities by subtracting suitable approximations in the corresponding singular limits,
\begin{eqnarray}
\label{eq:qqbartoggton_bold}
{\widehat {\cal M}}^{(2)}_{q\bar q \to gg \to n} \equiv {\cal M}^{(2)}_{q\bar q \to gg \to n}    
-  \Delta_1 \, {\cal M}^{(2)}_{q\bar q \to gg \to n}
-\Delta_2 \, {\cal M}^{(2)}_{q\bar q \to gg \to n} \, . 
\end{eqnarray}
Explicitly, we choose
\begin{eqnarray}
\label{eq:T1collphys-qqbar}
    \Delta_1 \, {\cal M}^{(2)}_{ q \bar q \to gg \to n} 
&=& -i\, g_s^2 \, \delta_{ij} \,  C_F \,
\frac{ \, {\bar  v}(p_2) \, \gamma^\beta u(p_1)\, p_1^\alpha}{p_1 \cdot p_2\,k^2\, k_1^2}
\nonumber \\ 
&& 
\hspace{-2.5cm}
 \times \sum_{{\rm perms}\, \{q_{n-1}\dots q_1\} } \left[
\frac{(2q_n-k_1)_\alpha}{ \left( k_1^2-2 k_1 \cdot q_n \right)}{\cal L}^{(n+1)}_\beta\left( l,k_2,q_n\dots q_1 \right) 
- (l\to l+k_1)
\right]
\end{eqnarray}
and 
\begin{eqnarray}
\label{eq:T2collphys-qqbar}
    \Delta_2 \, {\cal M}^{(2)}_{ q \bar q \to gg \to n} 
&=& i\, g_s^2 \, \delta_{ij} \,  C_F \, 
\frac{ {\bar  v}(p_2) \, \gamma^\alpha u(p_1)\, p_2^\beta}{p_1 \cdot p_2\,k^2\, k_2^2 }
\nonumber \\ 
&& 
\hspace{-2.5cm}
 \times \sum_{{\rm perms}\, \{q_{n-1}\dots q_1\} } \left[
\frac{(2q_n-k_2)_\beta}{ \left( k_2^2-2 k_2 \cdot q_n \right)}{\cal L}^{(n+1)}_\alpha\left( l,k_1,q_n\dots q_1 \right) 
- (l\to l+k_2)
\right].
\end{eqnarray}
As we have already remarked, the collinear counterterms vanish upon integration over the $l$ loop momentum, and the modified integrand ${\widehat {\cal M}}^{(2)}_{q\bar q \to gg \to n}$ of Eq.~\eqref{eq:qqbartoggton_bold}, which is free of collinear singularities locally, is equivalent to the original amplitude.

\section{Ultraviolet counterterms}
\label{sec:UVCT}

The infrared finite two-loop amplitude constructed above is still divergent from ultraviolet limits of integration, where linear combinations of the loop momenta $k,l$ become large. We can remove the ultraviolet singularities by subtracting suitable approximations. We follow a procedure that has also been explained in Refs.~\cite{Anastasiou:2020sdt,Anastasiou:2022eym}, highlighting here points which are particularly important for gluon fusion processes.

At one loop, the ultraviolet subtractions take the form, 
\begin{align}
\label{eq:Roper1L}
    \mathcal{M}_{n}^{(1),R} \equiv \mathcal{M}_{n}^{(1)} - \mathcal{R}_{l\to\infty}\left(\mathcal{M}_{n}^{(1)}\right) \, ,
\end{align}
where the second term approximates the integrand in the large loop momentum,  $l \to \infty$, limit. 
At two loops, we have 
\begin{align}
\label{eq:Mbold-uvfinite}
    {\cal H}_{n}^{(2),R} \cal \,\equiv& \, {\cal H}_{n}^{(2)} - \sum_{i} \mathcal{R}_{l_i(l,k) \to \infty}\left({\cal H}_{n}^{(2)}\right)  - \mathcal{R}_{k,l \to\infty} \left({\cal H}_{n}^{(2)}\right)
    \nonumber \\
    &
    + \mathcal{R}_{k,l \to\infty} \, \left(\sum_{i} \mathcal{R}_{l_i(l,k) \to \infty} \left({\cal H}_{n}^{(2)}\right)\right)\, ,
\end{align}
where ${\cal H}_{n}^{(2)}$ is defined in Eqs.~\eqref{eq:calH-threeterms} and \eqref{eq:GGsubtraction-termbyterm}.
The second term on the right-hand side approximates the integrand in limits where only a single loop momentum $l_i(l,k)$ at a time is taken to infinity, while a second independent loop momentum is kept fixed.
The third term on the right-hand side of Eq. \eqref{eq:Mbold-uvfinite} removes singularities when both independent loop momenta approach infinity (at the same rate). 
Finally, the last term removes the overlap of the previous two subtractions.

Similarly to Refs.~\cite{Anastasiou:2020sdt,Anastasiou:2022eym} and, previously, Ref.~\cite{Nagy:2003qn}, we formulate all UV counterterms for diagrams and infrared counterterms by performing a Taylor expansion of the integrand in the respective ultraviolet limit. We truncate the series, retaining only singular terms. The coefficients of the naive Taylor expansion in the ultraviolet limit are singular in the infrared. In order to avoid reintroducing infrared singularities with our ultraviolet counterterms, we further modify the Taylor expansion by giving a common mass $M$ into all propagator denominators~\cite{Nagy:2003qn}. While the ultraviolet counterterms and, in turn, the finite remainder depend on the auxiliary mass scale 
$M$, their sum remains independent of $M$.

As illustrative examples, we present the single ultraviolet counterterms that emerge in the second term on the right-hand side of Eq.~\eqref{eq:Mbold-uvfinite}. Additionally, we highlight the less intuitive appearance of ultraviolet counterterms on the right-hand side of Eq.~\eqref{eq:Roper1L}. The construction of the two-loop double ultraviolet counterterms in the last two terms of Eq.~\eqref{eq:Mbold-uvfinite} is relatively straightforward, and we will not comment further on them. 

Following the procedure of Refs.~\cite{Anastasiou:2020sdt,Anastasiou:2022eym}, the ultraviolet counterterm for propagator subgraphs with a general incoming fixed quark momentum $r$ reads, 
\begin{align}
    &
    \mathcal{R}_{k\to\infty}
    \includegraphics[width=0.165\textwidth, page=9,valign=c]{figures_standalone/tikz_UV_general.pdf}
 =  -g_s^2 \,C_F \, \delta_{ij}\,\left(
 \frac{(2-D)(\slashed{k}+\slashed{r})+ D m_q}{(k^2-M^2)^2} - \frac{(4-2D) k\cdot r\,\slashed{k}}{(k^2-M^2)^3}  \right)\, ,
 \label{eq:Rkself}
\end{align}
where $i$ and $j$ are the quark color indices. 
For one-loop vertex subgraphs with two fixed independent external momenta $r,q$  the ultraviolet counterterms read,
\begin{align}
    &
    \mathcal{R}_{k\to\infty}
    \includegraphics[width=0.2\textwidth, page=8,valign=c]{figures_standalone/tikz_UV_general.pdf}
    = -g_s^2\,Y_q \, C_F \, \delta_{ij}\frac{D}{(k^2-M^2)^2}, 
    \label{eq:RkHiggs}\\
&    \mathcal{R}_{k\to\infty}
    \includegraphics[width=0.3\textwidth, page=7,valign=c]{figures_standalone/tikz_UV_general.pdf}\nn\\
    &\hspace{1cm}= -g_s^3  \left(C_F-\frac{C_A}{2}\right) T_c\, \delta_{ij}\,  (2-D) \left( \frac{2k^{\mu}\slashed{k}}{(k^2-M^2)^3} - 
    \frac{\gamma^{\mu}}{(k^2-M^2)^2} \right) ,
    \label{eq:RgQED}\\
    &\nn\\
&    \mathcal{R}_{k\to\infty}
    \includegraphics[width=0.3375\textwidth, page=5,valign=c]{figures_standalone/tikz_UV_general.pdf}
    = 
    g_s^3 \, C_A\,T_c\, \delta_{ij}\, \frac{(2-D)k^{\mu}\;\slashed{k}}{(k^2-M^2)^3}, 
    \label{eq:Rkgss}
    \\
\label{eq:kplusluv}
    &
    \mathcal{R}_{k\to\infty}
    \includegraphics[width=0.3\textwidth, page=1,valign=c]{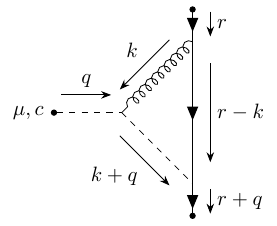}
    = -g_s^3\, \gamma^{\mu} \frac{C_A}{2} \, T_c\, \delta_{ij}\,  \frac{1}{(k^2-M^2)^2}.
\end{align}
In our assignment of momentum flows for the integrand of the amplitude, as we have defined them in the previous sections, the momentum $q$ always corresponds to the momentum of an external particle ($p_1$, $p_2$ or $q_i$ for $i=1,\ldots n$), and it is fixed to a finite value. The momentum $r$, which also needs to be kept fixed to finite values in deriving the above ultraviolet approximations, depends on the loop momentum flowing in the second loop. 

We note that the momentum $r$ can be a linear combination of both default loop momentum labels $k,l$, that are used in our definition of momentum flows for the amplitude integrand. An example diagram where this occurs is
\begin{equation*}
\includegraphics[width=0.35\textwidth, page=11 ,valign=c]{figures_standalone/tikz_sd_example.pdf} \, ,     
\end{equation*}
which, with momentum flows chosen as  in Eq.~\eqref{eq:momentumflow_B}, contains the triangle subgraph, 
\begin{align}
    \includegraphics[width=0.34\textwidth, page=10,valign=c]{figures_standalone/tikz_UV_general.pdf} 
\, 
.
\end{align}
Before taking the ultraviolet limit on the loop momentum $k$, the incoming momentum $l+k-p_1$ needs to be fixed to a finite value. The ultraviolet limit is taken by setting $l = -k+p_1+r$, where $r$ is fixed, followed by $ k\to \infty$ . This procedure yields the same UV approximation as in Eq. \eqref{eq:kplusluv}.

The form factor infrared counterterm of Eq.~\eqref{eq:scalarformfactor} is ultraviolet divergent. We remove this singularity with the following counterterm,
\begin{align}
    &\mathcal{R}_{k \to \infty}
         \includegraphics[width=0.21\textwidth, page=1,valign=c]{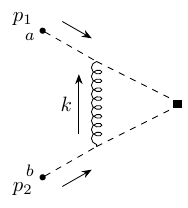}
    =  -i\,g_s^2 \,C_A \,\delta_{ab}\, \frac{1 }{(k^2-M^2)^2} \, .
    \label{eq:RopFormfactor}
\end{align}

As Eq.~\eqref{eq:Roper1L} indicates, ultraviolet singularities can also emerge in one-loop gluon-fusion amplitudes, at leading order in perturbation theory.
The one-loop integrand of $\widetilde{\mathcal{M}}_n^{(1)\mu\nu}$ also appears in subgraphs of the two-loop amplitude, as can be seen readily in the diagrams of Eq.~\eqref{eq:TwoLoopN}. 

Simple power counting shows that fermion loops with $n+2$ fermionic propagators have an ultraviolet degree of divergence equal to $1-n$ for an odd number of Higgs vertices $n$ and $2-n$ for even $n$.
Therefore, only the amplitudes ${\widetilde M}^{(1)}_n$ for $n=1,2$, corresponding to single and double Higgs production, are singular. Indeed, for fermion loops with a single external Higgs boson, we have that 
\begin{align}
\mathcal{R}_{l\to\infty}\, \widetilde{\mathcal{M}}_1^{(1)\,\alpha\beta} \delta_{ab}
&= 
\mathcal{R}_{l\to\infty} 
\left[ 
    \includegraphics[width=0.25\textwidth, page=1,valign=c]{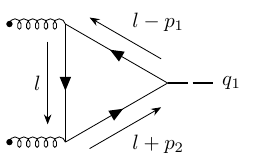} 
    +
    \includegraphics[width=0.27\textwidth, page=2,valign=c]{figures_standalone/tikz_oneloop.pdf}
\right]
\nonumber \\
&= -4\,g_s^2\,Y_q \,m_q\, \delta_{ab}\left( \frac{4\,l^{\alpha}l^{\beta}}{(l^2-M^2)^3} - \frac{ \eta^{\alpha\beta}}{(l^2-M^2)^2} \right).
    \label{eq:UVCTM1}
\end{align}
For the one-loop Higgs pair production amplitude, the ultraviolet counterterm is of the same form, 
\begin{align}
    \mathcal{R}_{l\to\infty} \, \widetilde{\mathcal{M}}_2^{(1)\,\alpha\beta}
    &= 
    -4\,g_s^2\,Y_q^2 \left( \frac{4\,l^{\alpha}l^{\beta}}{(l^2-M^2)^3} - \frac{ \eta^{\alpha\beta}}{(l^2-M^2)^2} \right). 
\label{eq:UVCTM2}
\end{align}
The local ultraviolet singularities of Eqs.~\eqref{eq:UVCTM1} - \eqref{eq:UVCTM2} do not yield an ultraviolet divergence, which is what is expected for Born amplitudes in renormalizable theories. Indeed, our local counterterm can be cast as a total derivative, vanishing upon integration within dimensional regularization, 
\begin{eqnarray}
\label{eq:evanCT}
    \int d^Dl \left( 4\frac{l^{\alpha}l^{\beta}}{(l^2-M^2)^3} - \frac{ \eta^{\alpha\beta}}{(l^2-M^2)^2} \right)
    =  \int d^Dl \,\frac{\partial}{\partial l_{\beta}} 
    \left(-\frac{l^{\alpha}}{(l^2-M^2)^2} \right) =0\,.
\end{eqnarray}

Finally, we should remark that we also encounter singular ultraviolet behavior in the infrared shift-counterterms of Eq.~\eqref{eq:T1collphys_diag1} and Eq.~\eqref{eq:T2collphys}.
Indeed, we have that 
\begin{eqnarray}
\label{eq:Rshiftgen}
&& 
\mathcal{R}_{ l\to \infty} \, 
\sum_{{\rm perms}\, \{q_{n-1}\dots q_1\} } 
 \left[
{\cal L}^{(n+1)}_\beta\left( l,k_2,\ldots \right) 
-{\cal L}^{(n+1)}_\beta\left( l+k_1,k_2,\ldots \right) 
\right] \nonumber \\ 
&& \hspace{2cm} 
= 
\mathcal{R}_{l\to\infty}\, k_1^{\alpha} \, \widetilde{\mathcal{M}}^{(1)}_{n\,\alpha\beta} =  k_1^{\alpha} \, 
\mathcal{R}_{l\to\infty}\,
\widetilde{\mathcal{M}}^{(1)}_{n\,\alpha\beta}  \, . 
\end{eqnarray}

\section{Analytic and numerical checks}
\label{sec:numcheck}

The formalism developed in this article is suited for deriving finite amplitude integrands with universal infrared counterterms for generic production processes of colorless final states via gluon fusion. 
In addition to presenting proofs of cancellations of local singularities in the integrand in the previous sections, we used our method to explicitly construct local infrared subtractions for the amplitudes of single-Higgs and Higgs-pair production. The generation of Feynman diagrams for the corresponding amplitudes and counterterms was carried out with QGRAF \cite{Nogueira:1991ex} and a custom Maple \cite{maple} code. Further manipulations were performed using FORM \cite{Vermaseren:2000nd, Tentyukov:2007mu, Kuipers:2012rf}. The evaluation of the integrands in various limits was performed using Maple. For the analytic integration of counterterms, we carried out standard reductions to master integrals using AIR~\cite{Anastasiou:2004vj}.

We verified that the analytic integration of our infrared and ultraviolet counterterms reproduces the divergences of the amplitude for single Higgs production, matching, for example, the results of Ref. \cite{Anastasiou:2020qzk}. Following an analogous procedure to Section 7 of Ref. \cite{Anastasiou:2022eym}, we also tested semi-numerically that all infrared singularities of the amplitudes in Higgs-pair production are indeed canceled locally, at the level of the integrand. This validated our method in an example process of high complexity, which involves a large number of two-loop diagrams.

\section{Conclusions}
Gluon fusion processes are primary mechanisms for producing Higgs and electroweak gauge bosons at hadron colliders. Signals from these processes are a central focus of ongoing and future experiments, as well as phenomenological studies aiming to test the Standard Model and its extensions in the Higgs sector (see, for example, Refs.~~\cite{ATLAS:2022hsp,CMS:2022dwd,ATLAS:2022qjq,ATLAS:2021hvg,Plehn:2005nk,Binoth:2006ym,Maltoni:2014eza,Papaefstathiou:2015paa,ATLAS:2022jtk}). 

Analytic computations of perturbative corrections to gluon-fusion cross-sections are challenging, since they typically involve two or more loops and a variety of massive particles, virtual and real. The presence of internal masses in particular hampers the analytic evaluation of these amplitudes. For example, the state-of-the-art computations of the NLO cross-section for the production of a Higgs boson pair have relied crucially on numerical methods~\cite{Borowka:2016ehy,Borowka:2016ypz,Davies:2019dfy,Baglio:2018lrj,Baglio:2020ini}, complemented by analytic results.

In this article, we have developed a novel framework that enables the numerical computation in momentum space of two-loop amplitudes for the production of a generic colorless final state in gluon fusion to a heavy quark loop. Our method is inspired by the infrared factorization of QCD amplitudes, building upon an approach introduced in Refs.~\cite{Anastasiou:2022eym,Anastasiou:2020sdt,Anastasiou:2018rib}.

Our factorization-based approach for constructing local infrared counterterms for two-loop gluon-fusion amplitudes presented similar advantages and challenges to those encountered in quark annihilation~\cite{Anastasiou:2022eym}. 
Notably, Ward identities ensure extensive cancellations within sets of diagrams in collinear limits, but are not complete. They typically result in pairs of integrals related by shifts of loop momenta, whose combined integrals cancel. The majority of our counterterms are of this kind, and their identification only requires consistent assignment of momentum flows among diagrams. Singular regions of this type were denoted as ``shift-integrable" above.

Crucial to the assignment of appropriate momentum flows and the combination of diagrams that lead to local factorization has been a decomposition of the tree triple-gluon vertex into three scalar-gauge boson interactions, as introduced in Sec.~\ref{sec:framework_notation} and applied in Sec.~\ref{sec:IRsing_channels}. Using the ``scalar decomposition", each infrared singular diagram can be divided in a truly factorizable part (Sec.~\ref{sec:factorization}), corresponding to collinear and soft singularities that survive in the amplitude, and a non-factorizable but ``shift-integrable" part, consisting of collinear singularities that cancel among diagrams (Sec.~\ref{sec:NonLocalSings}). Such contributions led to non-factorized differences of terms, known as ``shift mismatches'' in the quark-antiquark processes discussed in Ref.~\cite{Anastasiou:2022eym}. In Sec.~\ref{sec:one-loop_WI}, we recalled the standard diagrammatic demonstration of the one-loop Ward identity as it appears in the processes we study. This provided the necessary ingredients for the construction of  shift counterterms, which depended on our choices for the routing of loop momenta. The counterterms remove the non-local shift mismatches and integrate to zero, leaving the value of the amplitude after integration identical.
The factorizable part of the amplitude similarly required the cancellation of shift mismatches with a counterterm which integrates to zero. 

In summary, we found that the physical, factorized infrared singularities reside in graphs of a simpler ``scalar'' theory. Subsequently, we removed these singularities with a simple counterterm, specifically a $2\to 1$ one-loop form factor amplitude in the ``scalar'' theory times the Born amplitude. All short-distance information on the masses and momenta of the produced color-singlet particles remains in the hard function. We confirmed, as described in Sec.~\ref{sec:numcheck}, that these hard functions, regularized in the ultraviolet as described in Sec.~\ref{sec:UVCT}, are both locally integrable and convergent, which was our goal.

The local subtraction formula presented in Eq.~\eqref{eq:GGsubtraction}
provides finite integrands for generic gluon-fusion processes at two loops. 
In our construction, we have used the Feynman gauge, as it can be advantageous for the numerical evaluation of amplitudes. This gauge choice leads to simpler integrands with compact numerators and denominators that are quadratic in the loop momenta. In our framework, infrared counterterms are not applied to individual diagrams, as this could generate large cancellations in their sum due to gauge dependence. Instead, we construct a finite hard function by subtracting full gauge-invariant amplitudes.

These finite integrands still contain integrable singularities that prevent their direct numerical integration. Threshold singularities, in particular, are more transparent in representations of the amplitudes in which the energy components of loop momenta are integrated out~\cite{Capatti:2022mly,Capatti:2023shz,Sterman:2023xdj,Catani:2008xa,Bierenbaum:2010cy,Runkel:2019yrs,Capatti:2019ypt,Sterman:1993hfp,Sterman:1995fz,Soper:1999xk}.
Recently, it was demonstrated that novel subtractions of threshold singularities~\cite{Kermanschah:2021wbk,Kermanschah:2024utt,Kermanschah:2024jbp} are compatible with the infrared subtractions introduced in Refs.~\cite{Anastasiou:2020sdt,Anastasiou:2022eym}. We believe that the subtractions presented in this article, complemented by the threshold subtraction method of Refs.~\cite{Kermanschah:2021wbk,Kermanschah:2024utt,Kermanschah:2024jbp}, will lead to practical numerical evaluations of the gluon fusion amplitudes as well. 

Furthermore, we believe that the techniques we introduced, including the scalar decomposition, momentum flow assignments, and shift counterterms of Section~\ref{sec:NonLocalSings}, will serve as essential components for extending the local factorization formalism to more complex processes involving gluons. We look forward to future developments in this direction.

\section*{Acknowledgements}
We would like to thank Yao Ma for useful discussions. This work was supported by the National Science Foundation, grant PHY-2210533.

%\newpage
\bibliographystyle{JHEP}
\bibliography{biblio}

\end{document}